\pdfoutput=1
\documentclass[preprint2]{aastex}
\usepackage{natbib}
\usepackage{amssymb}
\usepackage{multirow}

\newcommand{\hone}{{{\sc H i}~}}

\slugcomment{Draft, \today}

\shorttitle{\hone in the M81 Filament}
\shortauthors{Chynoweth et al.}

\begin{document}

\title{\hone Clouds in the M81 Filament as Dark Matter Minihalos--A Phase-Space Mismatch.}

\author{Katie M. Chynoweth}  
\affil{Naval Research Laboratory,
Washington, DC 20375}
\author{Glen I. Langston}
\affil{National Radio Astronomy Observatory,
Green Bank, WV 24944}
\author{Kelly Holley-Bockelmann}  
\affil{Vanderbilt University, Physics and Astronomy Department, 1807 Station B, 
Nashville, TN 37235}

\begin{abstract}

Cosmological galaxy formation models predict the existence of dark matter minihalos surrounding galaxies and in filaments connecting groups of galaxies.   The more massive of these minihalos are 
predicted to host \hone gas that should be detectable by current radio telescopes such as the GBT. 
We observed the 
region including the M81/M82 and NGC 2403 galaxy groups, searching for observational evidence of an \hone component associated with dark matter halos within
the ``M81 Filament", using the Robert C. Byrd Green 
Bank Telescope (GBT). 
The map covers an 8.7$^{\circ}$ $\times$ 21.3 $^{\circ}$ (480 kpc $\times$ 1.2 Mpc) region centered between the M81/M82 and NGC 2403 galaxy groups. 
Our observations cover a wide velocity range, from -890 to 1320 km s$^{-1}$,
which spans much of the range predicted by cosmological N-body simulations for dark matter minihalo velocities. 
Our search is not complete in the velocity range -210 to 85 km s$^{-1}$,
containing Galactic emission and the HVC Complex A.
For an \hone cloud at the distance of M81, with a size $\leq 10$ kpc, 
our average 5$\sigma$ mass 
detection limit is 3.2 $\times$ 10$^{6}$ M$_{\odot}$, for a linewidth of 20 km s$^{-1}$.
We compare our observations to two large cosmological N-body simulations
and find that the simulation predicts a significantly greater number of
detectable minihalos than are found in our observations, and that the simulated minihalos do not match the phase space of observed \hone clouds.
These results place strong constraints on the \hone gas that can be associated 
with dark-matter halos. Our observations indicate that the majority of extragalactic 
\hone clouds with a mass greater than 10$^{6}$ M$_{\odot}$ are likely to be generated through cold accretion, or via tidal stripping caused by galaxy interactions.

\end{abstract}


\section{Introduction}

$\Lambda$CDM cosmological models (e.g., \citet{dek86, kly99}) for structure formation in the 
universe predict many compact dark-matter structures, {\it minihalos}, 
in the regions near galaxies and galaxy groups.  \citet{ric09} predicts that minihalos may have a late phase of
gas accretion, which will be detectable in \hone observations. It has been proposed that at least part of the population of neutral hydrogen High Velocity Clouds (HVCs) surrounding the Milky Way may represent these \hone-embedded dark matter halos \citep{blitz99,braunburt99,gio10}. Other possible origins of the HVCs include galactic fountains \citep{bre80}, tidal stripping from major and satellite galaxy interactions \citep{put03,chy08},  and cold accretion of primordial gas \citep{Dusan:2009}.
Observations of the north Galactic polar region by the ALFALFA survey may have revealed low \hone mass minihalos surrounding the Milky Way \citep{gio10}; however, these clouds may also be part of the Magellanic Stream.
Searches for \hone counterparts to these minihalos within galaxy groups have yielded non-detections or detections that do not match the predicted phase space
(e.g. \citet{ pis07, chy08, chy09}).

In this paper, we conduct an \hone survey of a 8.7$^{\circ}$ $\times$ 21.3 $^{\circ}$ region of a nearby complex of galaxies known as the ``M81 Filament" \citep{kar02}, in an extensive search for \hone clouds that
may trace in \hone a population of ~$10^8-10^{10} M_\odot$ dark matter minihalos~\citep{krav04}. The M81 Filament includes the M81/M82 and NGC 2403 galaxy groups and the region between them.  It is likely that the linearly shaped set of galaxies is contained within a dark-matter filament of the type commonly seen to form in cosmological simulations.
The extent of the dark-matter and hot gaseous halos of these two groups is unknown -- in fact, the M81/M82 group 
is sometimes considered to encompass the NGC 2403 group as part of one cluster \citep{beg06, kar07}.  If this is true, then we may expect to detect pressure-confined clouds that trace cold accretion and \hone-laced dark matter halos in between the two groups, as well as galaxy interaction-generated \hone clouds near the strongly interacting M81 Group. 
With wide velocity coverage and high sensitivity over such a large area, this campaign was designed create a census of the more massive \hone clouds, which may be
counterparts to minihalos or generated by galaxy interactions.

In Section \ref{obsec}, we summarize the observations. In Section \ref{datasec}, we describe the data reduction process. In Section \ref{interp}, we discuss the interpretation of the data and sensitivity limits. Section \ref{ressec} describes our detections. In Section \ref{simsec}, we compare our results with numerical simulations. Finally, Section \ref{consec} contains discussion, and we draw conclusions and discuss future work. In Appendix A, we present the \hone properties of the galaxies in the observed volume.

\section{Observations}
\label{obsec}
\begin{figure}
\centering
\includegraphics[width=5in]{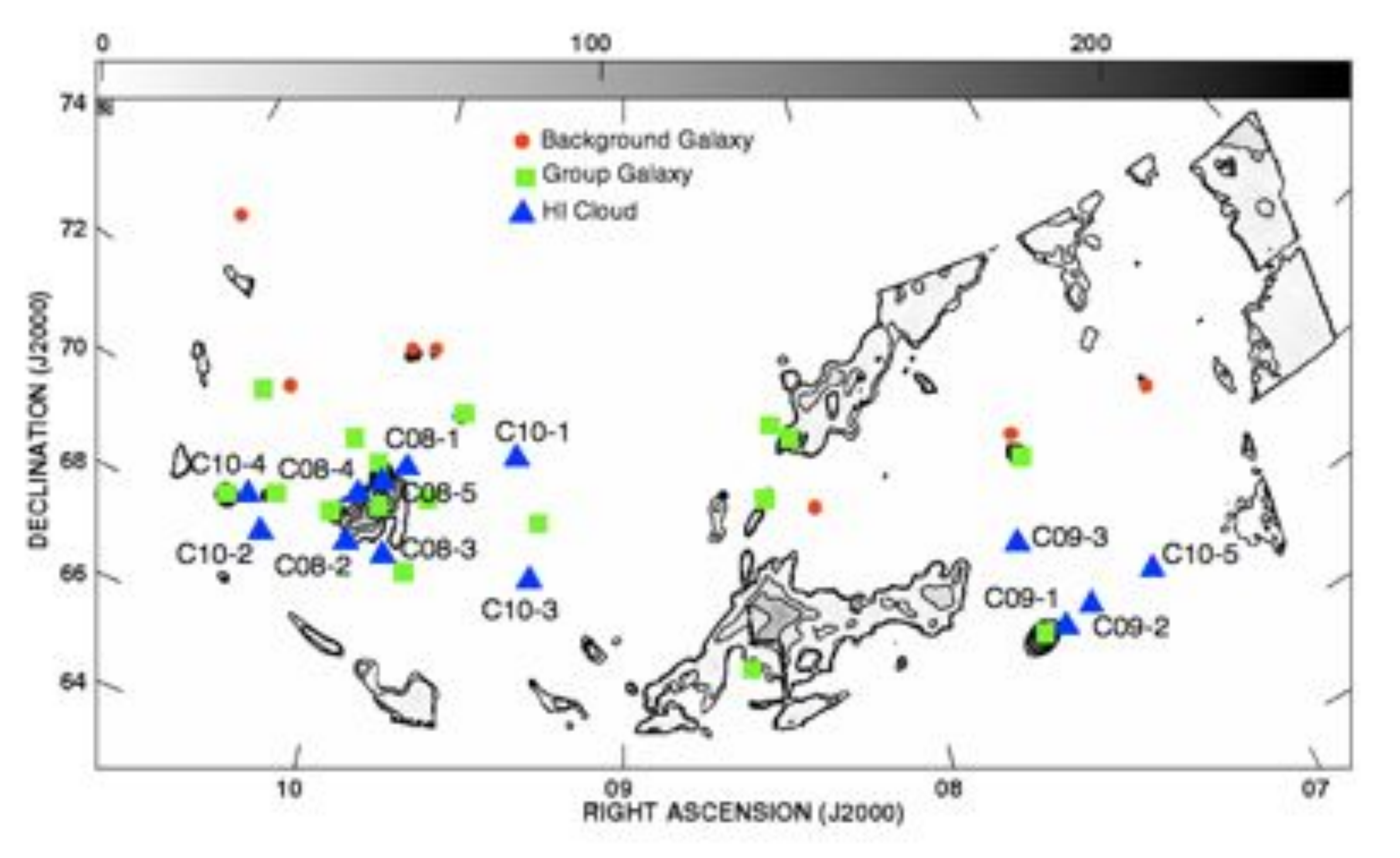}
\caption{\hone column density map of the observed area. Background galaxies, group galaxies, and \hone clouds are marked. The diagonal filament of \hone through the middle of the map is Milky Way HVC Complex A.} 
\label{mom0}
\end{figure}

Observations were carried out with the 100m Robert C. Byrd Green Bank Telescope (GBT) of the 
NRAO\footnote{The National Radio Astronomy Observatory (NRAO) is a facility of the 
National Science Foundation operated under cooperative agreement by Associated Universities, Inc.}
in 40 sessions between June 2003 and August 2009.  This includes data taken for previously published observations of the M81/M82 and NGC 2403 galaxy groups; see \citet{chy08} and \citet{chy09} for details. We combined the 40 observing
sessions into a single 8.7$^{\circ}$ $\times$ 21.3 $^{\circ}$  map centered at 08$^h$40$^m$37.0$^s$ , 
69$^{\circ}$17$^`$16$^{``}$. Figure
\ref{mom0} shows a layout of the
observations. 

\begin{figure}
\centering
\includegraphics[width=5in]{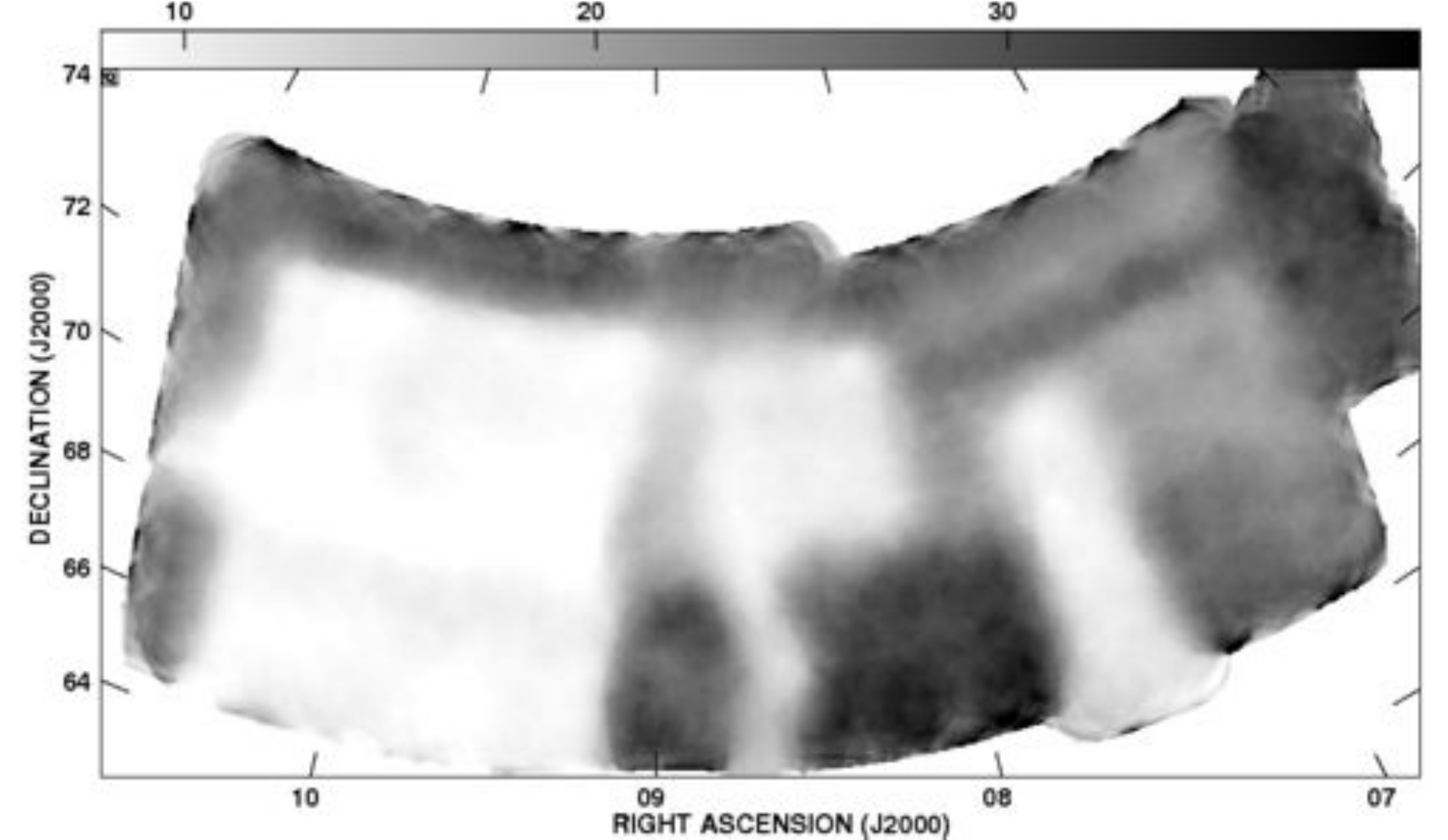}
\caption{Sensitivity of the mapped area. The greyscale ranges from 8 to 40 mJy beam$^{-1}$, corresponding to mass sensitivity of 2.5 to 12.9 $\times$ 10$^5$ M$_{\odot}$. The colorbar above the figure is in units of mJy beam$^{-1}$.}
\label{RMSfig}
\end{figure}

The filament was observed by moving the telescope in declination and sampling every $3\arcmin$ with an integration 
time of 1-3 seconds per sample, depending on the observation session.  Strips of constant declination were 
spaced by  $3\arcmin$.  Corresponding maps were made by moving the telescope in right ascension to form a 
`basket weave' pattern over the region. 
Total integration time for the entire map was approximately 187 hours.  In Figure \ref{RMSfig} it is possible to infer the RMS noise and corresponding mass sensitivity of the map as a function of position. 

The GBT spectrometer was used with a bandwith of either 12.5 or 50 MHz, depending on the observation session. The 
combined bandwidth for the final map is 10.5 MHz, corresponding to a velocity range from -890 to 1320 km s$^{-1}$. The typical system temperature for each channel of the dual-polarization 
receiver was $\sim$ 20 K.

Our observations cover a large angular area in order to make a complete study of 
the \hone properties across the entire extent of the filament. The beam size of the GBT in the 21 cm line is
$9\arcmin$, which maps to approximately 10 kpc at the distance of the M81 filament. This is well 
matched to the angular size of known \hone clouds above our mass detection
threshold \citep{thilker04, chy08}. Predictions for the sizes of dark-matter minihalos range from 3-30 kpc in diameter \citep{putmoore02}; we may be unable to detect minihalo clouds with a $\le$ 10 kpc diameter due to beam dilution.

\section{Data Reduction}
\label{datasec}

The GBT data were reduced in the standard manner using the GBTIDL and AIPS\footnote{Developed by the National Radio Astronomy Observatory;
documentation at http://gbtidl.sourceforge.net, http://www.aoc.nrao.edu/aips} 
data reduction packages.

In order to match our velocity resolution to the expected linewidths of \hone 
clouds in the group, spectra were smoothed  
to a channel spacing of 24.4 kHz, corresponding to a velocity resolution of 
5.2 km s$^{-1}$. A reference spectrum for each of the observation sessions 
was made using an 
observation of an emission-free region, usually from the edges of the maps. The reference spectrum was then used to 
perform a (signal-reference)/reference calibration of each pixel. 
The calibrated spectra were scaled by the system temperature, corrected 
for atmospheric opacity and GBT efficiency. We adopted the GBT efficiency equation 
(1) from \citet{lang07} with a zenith atmospheric opacity 
$\tau_{0}$ = 0.009. Velocities are in the kinematic LSR 
velocity frame. 

The frequency range observed was relatively free of RFI, with less than 0.2\% of 
all spectra adversely affected. The spectra exhibiting RFI were 
identified by tabulating the RMS noise level in channels free of neutral 
hydrogen emission.  Spectra that showed high values of RMS
noise across many channels were flagged and removed.  
Observations were gridded using the AIPS task SDIMG, which also averages 
polarizations. After amplitude calibration and gridding, a 1st-order polynomial 
was fit to line-free regions of the spectra and subtracted from 
the gridded spectra using the AIPS task IMLIN. Only the channels in the velocity range -400 to -800 km s$^{-1}$ were 
used for the fit. This simple baseline fit was extrapolated to the positive velocity range, 
where galaxies make a baseline fit unreliable. This baseline velocity range for the fit yielded a flat baseline for areas free of strong continuum radio sources, which is true of the majority 
of the region.
The effective angular resolution, determined 
from maps of 3C286, is 9.15\arcmin $\pm$ 0.05\arcmin. 
To convert to units of flux density, we observed the calibration source 3C286, 
whose flux density is 14.57 $\pm$ 0.94 Jy at 1.418 GHz  
\citep{ott94}. The calibration from 
K to Jy was derived by mapping 3C286 in the same way that the
 \hone maps were produced. 
After all corrections for the GBT efficiency and the mapping process, the 
scale factor from K/Beam to Jy/Beam images is 0.43 $\pm$ 0.03. Due to the patchwork nature of the observations, the RMS noise varies considerably across the datacube, ranging between 6-30 mJy/beam.
The average RMS noise in the final data cube is 
 20 mJy per 24.4 kHz channel.  Figure \ref{RMSfig} shows the map sensitivity as a function of position. The instrumental parameters are 
summarized in Table 2, and Table 3 gives a summary of the observations.  The spectral line images are available on-line\footnote{ http://www.nrao.edu/astrores/m81/ }.

\section{Data Interpretation}
\label{interp}

In order to detect either \hone clouds associated with dark matter minihalos , or cold accretion, a large-area map is required. \hone clouds in dark matter minihalos are predicted to be found at distances of up to 1 Mpc from a major halo \citep{blitz99}. Cold accretion clouds and interaction-generated clouds are expected to be contained within a $\sim$ 60 kpc radius around large galaxies \citep{chy08, Dusan:2009}.  Our map has a projected size of 480 kpc $\times$ 1.2 Mpc, which is adequate to detect all three phenomena.
 
Spectral maps were made in the velocity range overlapped by all observations, from -890 to 1320 km s$^{-1}$.  Spectral maps were visually inspected for possible new \hone clouds. We required a cloud candidate to be visible in at least two channels, therefore
the lowest velocity width that could be detected was 10 km s$^{-1}$ .  For each candidate, a plot of 
intensity versus velocity was also produced and checked. 
At the locations 
of strong continuum radio sources, the linear spectral baseline fit was occasionally 
poor. In these cases, the spectra would show a broad slope, $>$ 400 km s$^{-1}$ wide, either 
increasing or decreasing with frequency. Candidates in the regions with poor spectral baseline were also discarded. Only 26 of the more than 10,000 GBT beams covering the map showed spectra with poor spectral 
baselines. 

The mass detection threshold was
calculated assuming that clouds would be unresolved in the GBT beam, using the
relation:

\begin{equation}
\left(\frac{\sigma_{M}}{M_{\odot}}\right)=2.36 \times 10^5 \left(\frac{D}{{\rm Mpc}}\right)^2 \left(\frac{\sigma_{s}}{{\rm Jy} }\right)
\left(\frac{\Delta V}{{\rm km}\hskip 1 mm {\rm s}^{-1}}\right) \sqrt N ,
\end{equation}

\noindent where $D$ is the average distance to the filament, $\sigma_s$ is the RMS noise in one channel, $\Delta V$
is the channel width, and $N$ is the number of channels required for a secure detection. For this calculation, we used an \hone cloud linewidth of 
20 km s$^{-1}$, corresponding to $N=4$. This assumed linewidth strikes a balance between narrower HVC lines \citep{wvw97} and the wider linewidths of massive \hone clouds \citep{chy08}.

Adopting the M81 distance of 3.6 Mpc, the 1$\sigma$ mass detection threshold ranges from 2.5 to 12.9 $\times$ 10$^5$ M$_{\odot}$, as shown in Figure \ref{RMSfig}. The average 5$\sigma$ detection threshold for \hone mass was  
3.2 $\times$ 10$^{6}$ M$_{\odot}$. With this threshold, we would be able to detect analogues to
the most massive clouds around M31 and M33, the M81/M82 clouds, the M101 cloud, 
and large Milky Way objects such as Complexes C and H and the Magellanic Stream. We would also be able to detect cold accretion clouds found in the simulations of \citet{Dusan:2009}.

\section{New Identifications} 
\label{ressec}

We discovered 5 new \hone clouds.   We divide these clouds
into two groups,  those associated with the M81 Filament and those associated with
the Milky Way. We determine the most likely association of each cloud based on its position and velocity. There is some overlap in velocity space between the M81 Filament and the 
Milky Way, and over the velocity range -85 to 25 km s$^{-1}$, we can not discriminate between local 
and distant emission. In addition, the Galactic HVC Complex A, which ranges in velocity from -120 km s$^{-1}$ to -200  km s$^{-1}$, coincides with part of the mapped region \citep{wvw97}.  Therefore we cannot distinguish \hone clouds in this position and velocity range from Complex A.

\begin{figure}
\centering
\includegraphics[width=5in]{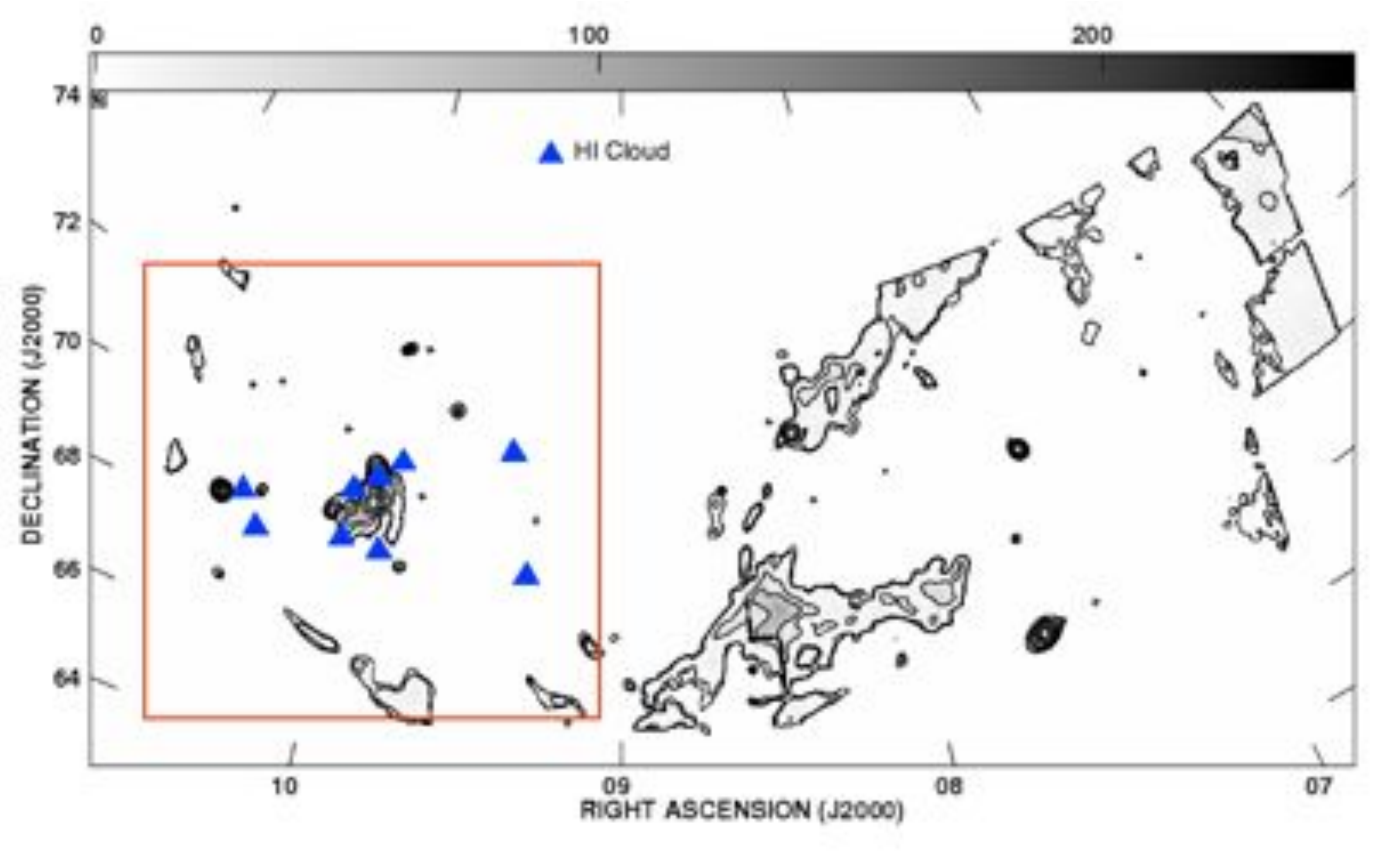}
\caption{ \hone clouds that are located within the M81 Filament are labeled with triangles.The region of the M81 Filament that was compared to a cosmological dark-matter simulation is indicated with a red box.}
\label{fil_analysis}
\end{figure}

\begin{figure}
\centering
\begin{tabular}{cc}
\includegraphics[width=3in]{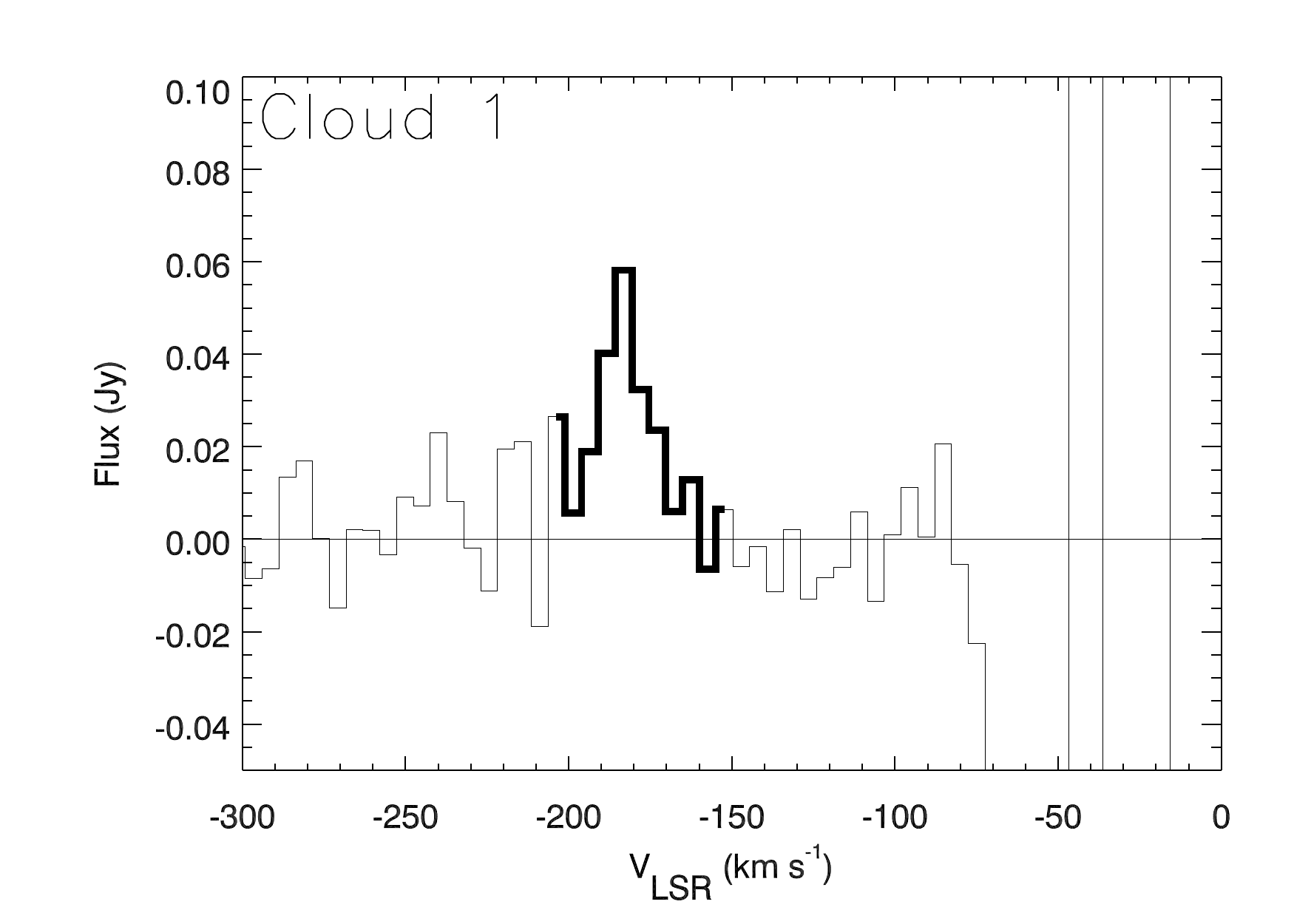}  &
\includegraphics[width=3in]{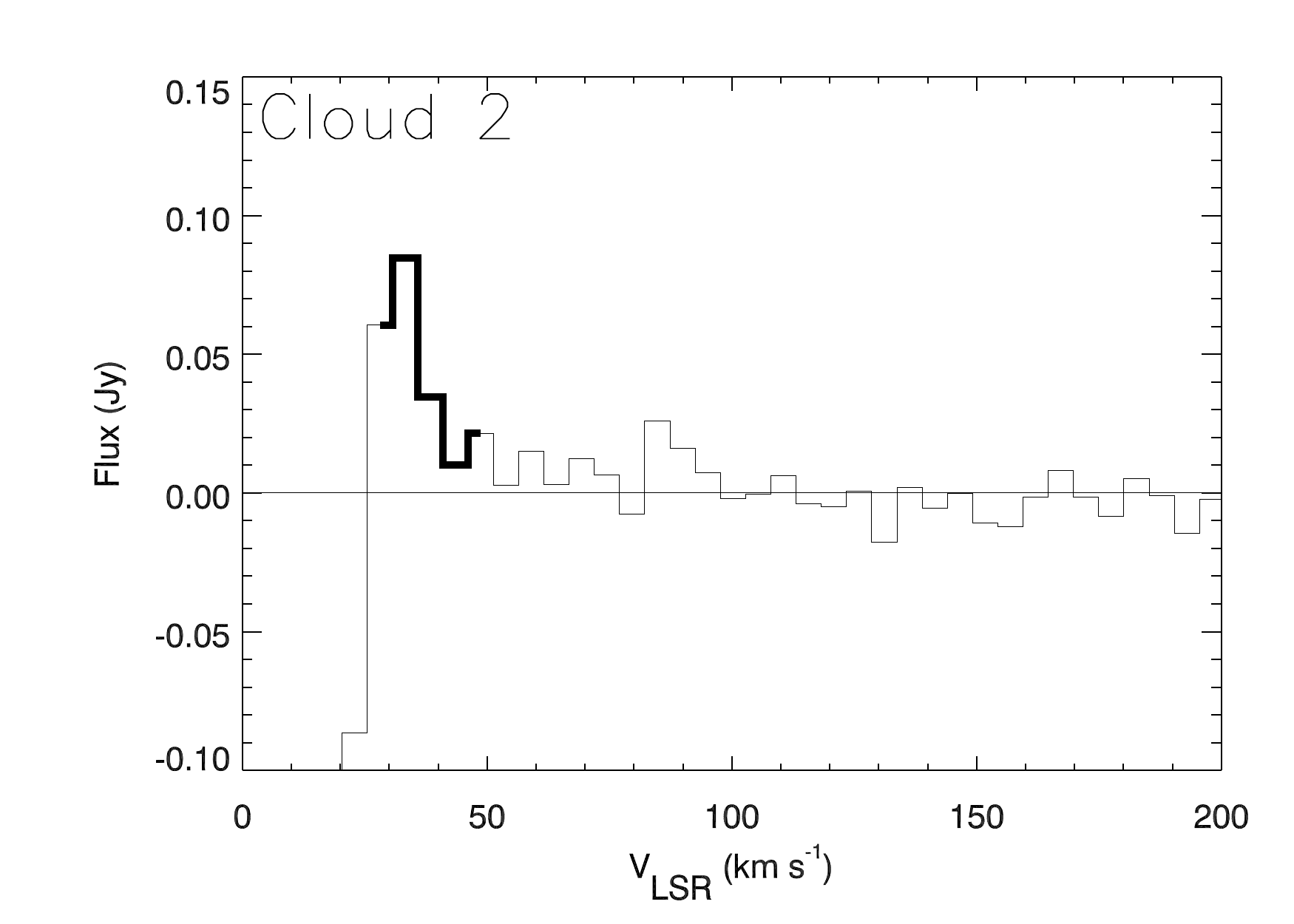}  \\
\includegraphics[width=3in]{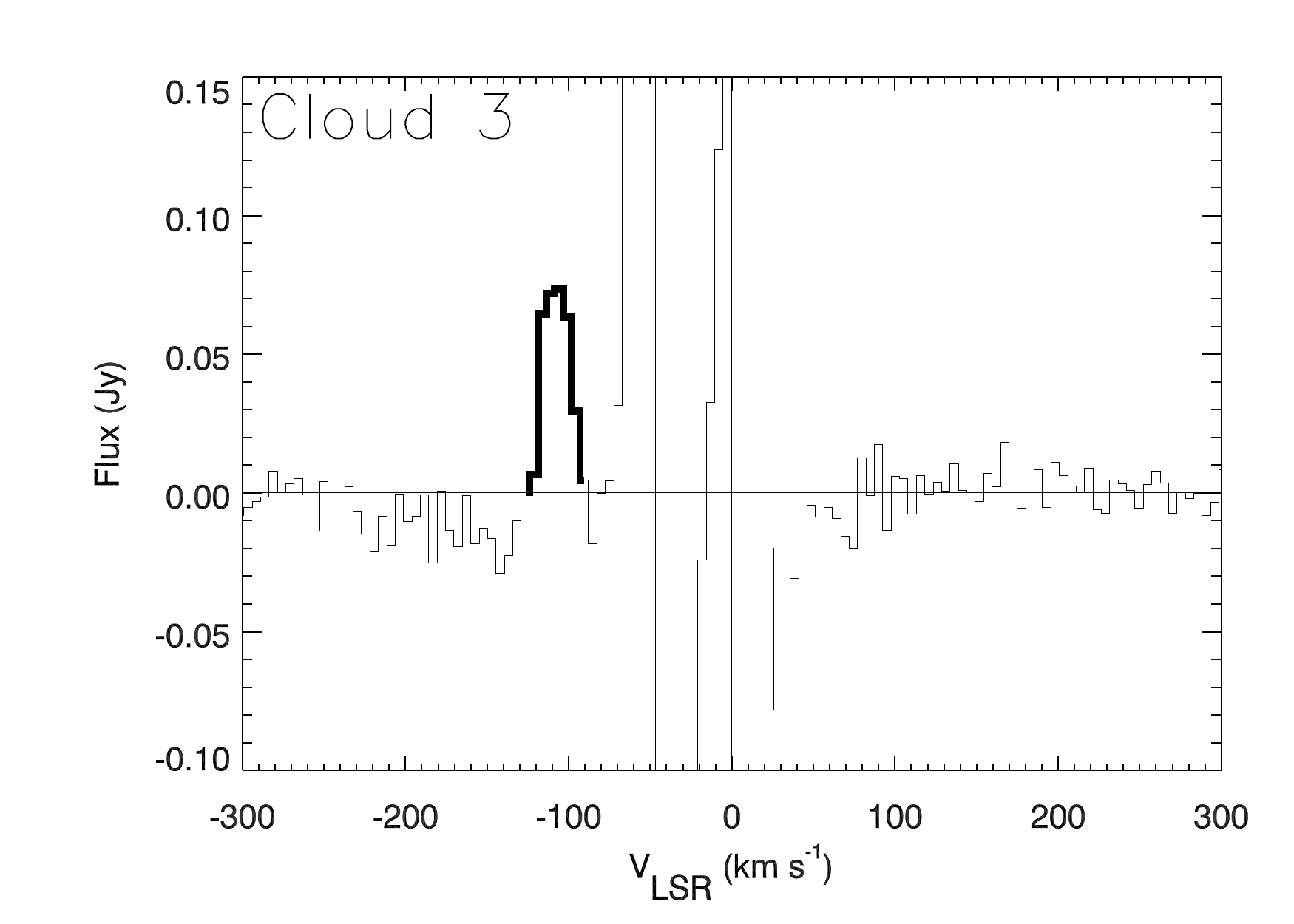}  & 
\includegraphics[width=3in]{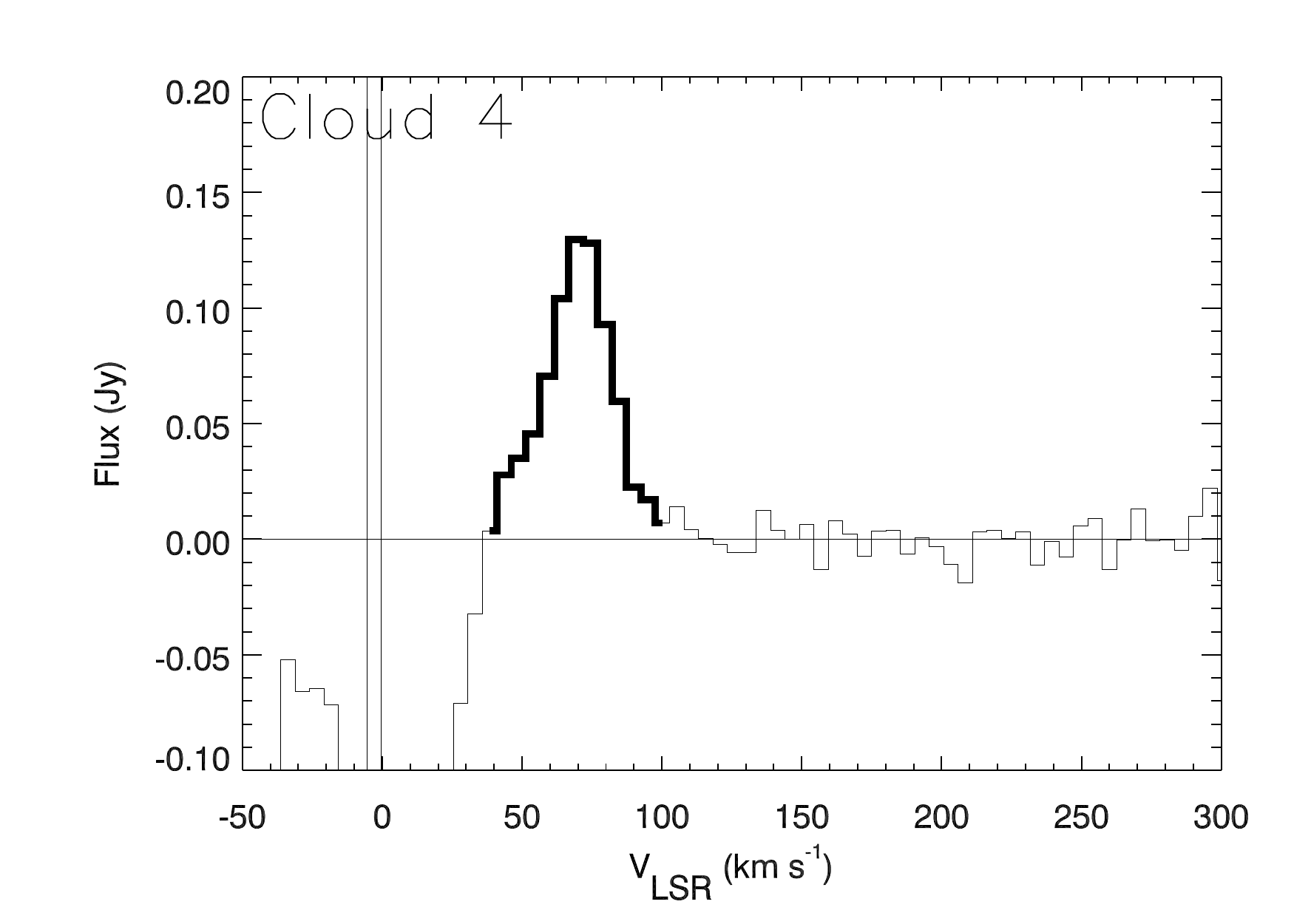} \\
\includegraphics[width=3in]{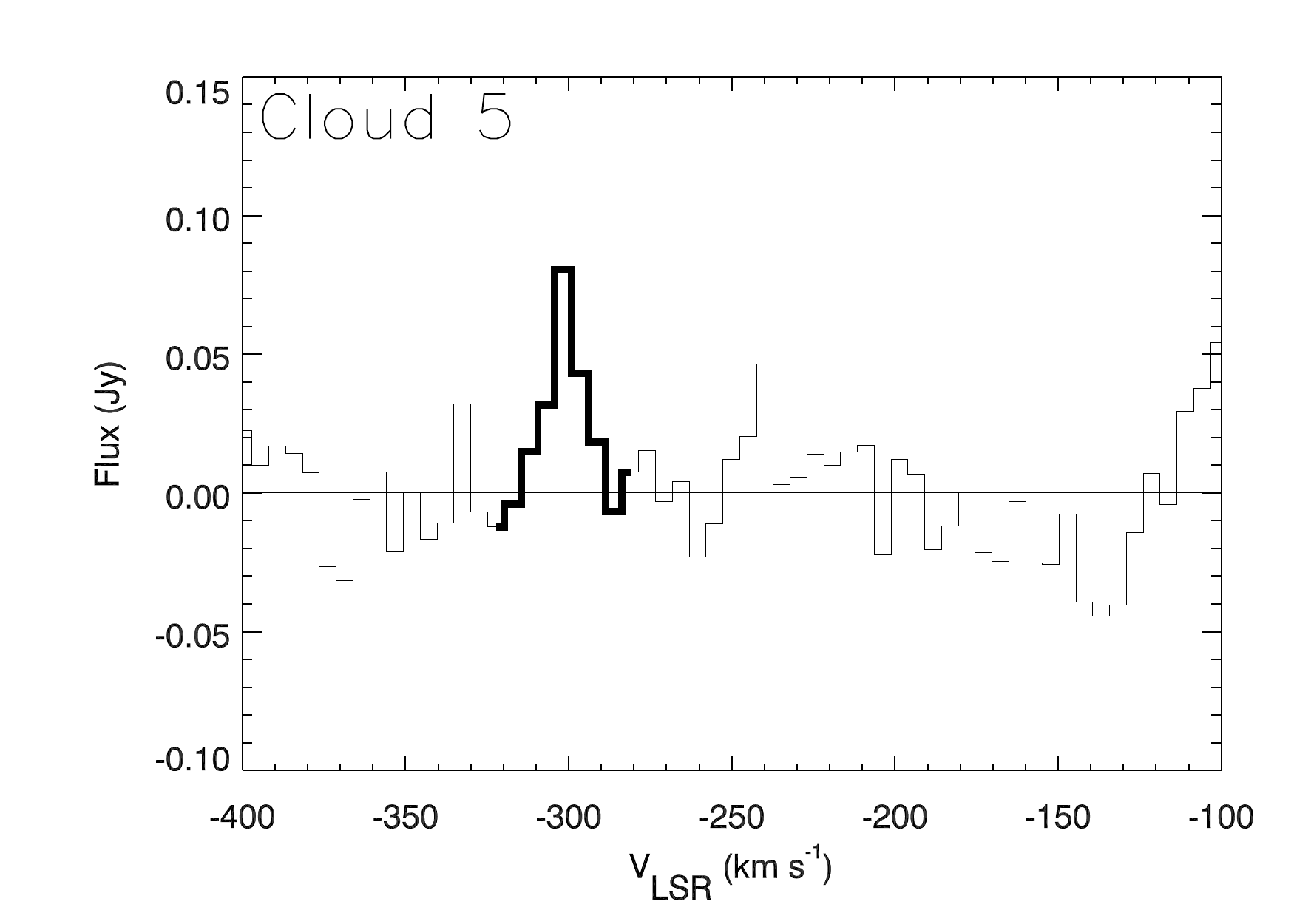} & \\
\end{tabular}
\caption{\hone intensity profiles of new \hone clouds. The regions of the
spectrum used to calculate \hone mass are highlighted in bold.}
\label{cloud_ispec}
\end{figure}

All of the \hone clouds we previously detected  \citep{chy08, chy09} are
visible in this extended image.
Their locations are marked in Figure \ref{mom0}. In order to unify the clouds identified in this paper with those in \citet{chy08} and \citet{chy09}, and any \hone clouds identified with the GBT in our current and future observations, we have re-designated the clouds with the identifier GBC for ``Green Bank Cloud", and the coordinates of the cloud, i.~e.~ GBC Jhhmmss.s+ddmmss. These designations are found in Table \ref{fil_cloudprop}.
To distinguish the newly discovered clouds from the clouds previously discovered,
the new clouds are designated with C10-1 to C10-5. 
Table \ref{fil_cloudprop} lists their properties. Spectra are shown in Figure \ref{cloud_ispec}.  Figure \ref{fil_analysis} shows the locations of all \hone clouds that we have determined to be located within M81 Filament.

Cloud masses were calculated using

\begin{equation}
\left(\frac{M}{M_{\odot}}\right)=2.36 \times 10^5 \left(\frac{D}{{\rm Mpc}}\right)^2 \left(\frac{F}{{\rm Jy}}\right)
\left(\frac{\Delta V}{{\rm km}\hskip 1mm {\rm s}^{-1}}\right),
\end{equation}

\noindent where $F$ is the total flux of the cloud. Error estimates on the masses difficult to obtain, given the wide range in RMS 
noise values over the map. The dominant error 
contribution is the approximately 7\% uncertainty in the absolute calibration for these observations.

\subsection{M81 Filament Clouds}

Cloud C10-1 (GBC J092635.8+702850) is to the northwest of the M81 group. With a velocity of -178 km s$^{-1}$, Cloud C10-1 may be part of Complex A. However, Complex A does not overlap in position with Cloud C10-1, so we include this cloud in further analysis of \hone clouds in the M81 Filament.

Cloud C10-2 (GBC J101926.3+675222) is south of IC 2574 and an adjacent HIJASS 
source, HIJASS J1021+6840 \citep{boy01} and has a velocity of 38 km s$^{-1}$.   
Cloud C10-3 (GBC J091952.7+680937) is to the southwest of the M81 group and has a velocity of -108 km s$^{-1}$. 
Cloud C10-4 (GBC J1022:39.1+684057) is located near the previously reported HIJASS source. This  cloud is relatively bright and is a few arc-minutes east of the HIJASS source.  
\citet{boy01} find the HIJASS source is extended in the direction of IC 2574.   
Our observations have higher angular resolution, and in these observations Cloud C10-4 is clearly
separated from HIJASS J1021+6840.  Cloud C10-4 has a velocity of 69 km s$^{-1}.$ 

\subsection{Milky Way Clouds}

Cloud C10-5 (GBC J071051.4+654428)  is west of the clouds reported in \citet{chy09} and shows a high negative velocity. It is visible in 5 channels and is extended 
spatially.  It is offset by about 100 km s$^{-1}$ from Complex A, and may be a separate Milky Way HVC. Cloud C10-5 has a velocity -301 km s$^{-1}$ , which falls outside the 
velocity range of Complex A \citep{wvw97}.   There are Galactic HVCs detected with velocities a large as -450 km s$^{-1}$ \citep{wvw91}.
These observations can not confidently determine whether Cloud 10-5 is associated with the M81 Filament or with the Milky Way.

\section{Comparison with Numerical Simulations}
\label{simsec}

In order to determine whether the \hone clouds in the M81 Filament have the properties expected for gas-embedded dark-matter halos, we have compared their properties statistically with the properties of dark matter halos generated by N-body simulations. Analysis includes the clouds detected in this paper, in addition to the clouds from \citet{chy08} and \citet{chy09}. For this analysis we only include the \hone clouds located in the 8$^{\circ}$ $\times$ 
8$^{\circ}$ region centered on M81, due to the greater RMS noise and contamination by Complex A in other regions of the map. Therefore this analysis excludes clouds C10-5, C09 -1, C09-2, and C09-3. Figure \ref{fil_analysis} shows the region of the M81 Filament that was compared to the simulation, with \hone clouds marked.

\begin{figure}
\centering
\includegraphics[width=5in]{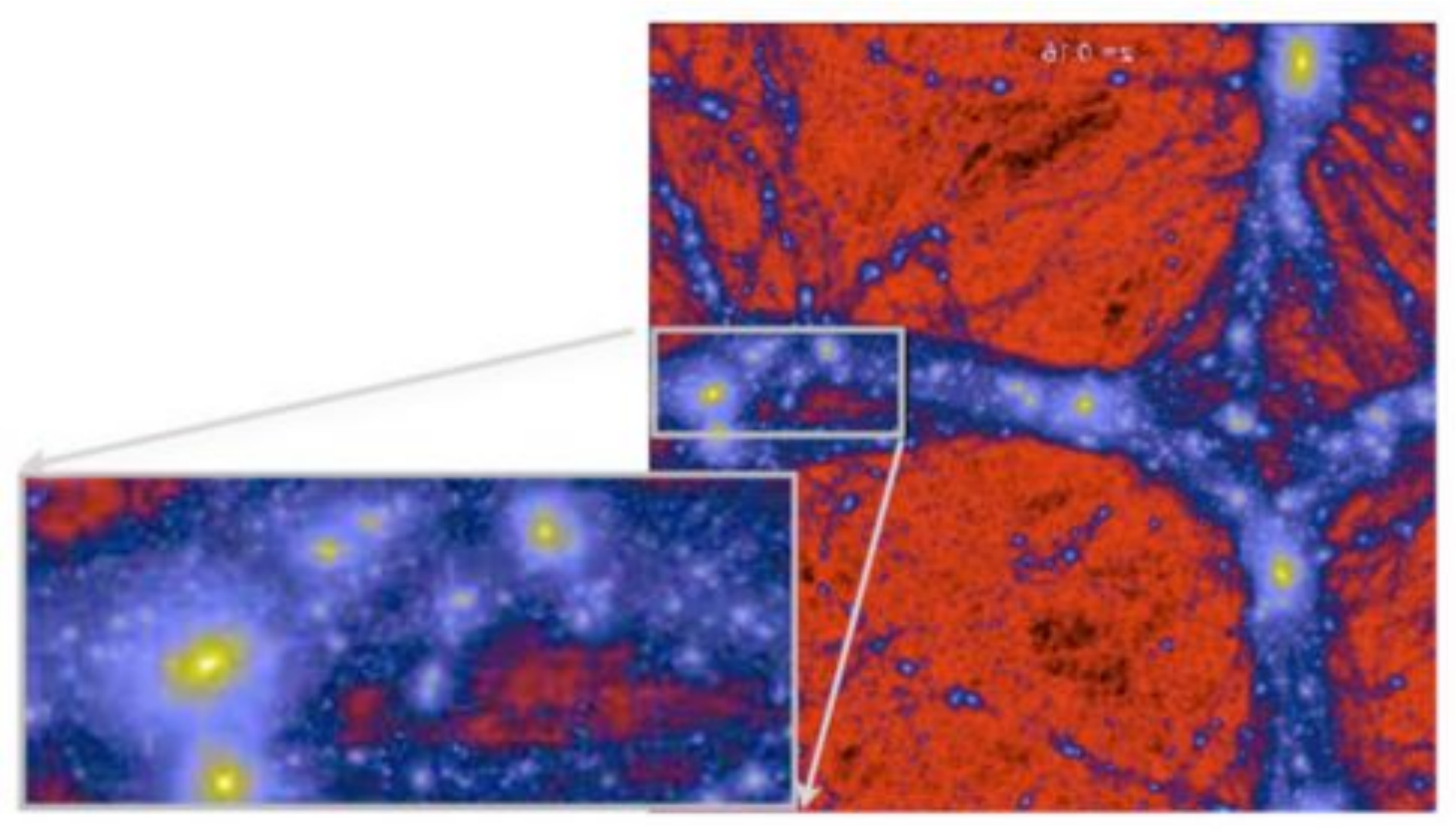}
\caption{Snapshot of the projected density of dark matter in the cosmological simulation used to find expected properties of halos. The large box is 10 Mpc on a side. The highlighted box corresponds to the angular size of our observations.}
\label{simfig}
\end{figure}

We conducted two dark matter-only cosmological N-body simulations for comparison. One simulation uses initial conditions from WMAP3, and one from WMAP5. We simulated a 50$^3$ Mpc$^3$ volume of the universe from z=149 to z=0 using 256$^3$
particles. At redshift zero, we selected a
volume 10 Mpc on a side to resimulate with 512$^3$ particles with a
'zoom' technique aimed to preserve the tidal field outside the higher
resolution volume. This smaller volume was selected to host a
$\sim 10^{12} M_\odot$ halo at z=0 -- this halo is consistent with the M81
galaxy. Figure \ref{simfig} shows a z=0 snapshot of the simulation.  We used a friends-of-friends algorithm with a linking length of $b=0.2$ to
identify dark matter halos, and used a SUBFIND technique to determine the bound subhalos.

In order to connect the numerical simulations with the properties of \hone observations, 
we require an estimate of the \hone to dark matter mass fraction. The distribution of \hone gas compared to the dark matter halo distribution is very uncertain. For the purpose of comparing our observations with the \hone clouds, we assume that \hone gas follows the dark matter distribution. We include in our comparison the prescription of \citet{gne00}, which includes a mass cutoff such that a dark matter halo with mass less than 2 $\times$ 10$^8$ M$_\odot$ has no associated \hone gas. We take as the scaling factor from dark matter to \hone gas mass 
the \citet{zwaan03} results, summarized in \citet{fuku04}. 
Their values yield a \hone to dark matter 
mass fraction of 0.0018 $\pm$ 0.0003. 

After scaling the simulated dark matter to \hone, the WMAP3 simulation contains 7 dark matter halos with a simulated \hone mass between 10$^{9-10}$ M$_{\odot}$, and the WMAP5 simulation contains 8.  This is approximately consistent with the M81 \hone mass of 2.67 $\times$ 10$^{9}$ M$_{\odot}$. For each of these 'major' halos, we extract a box corresponding to the 8$^{\circ}$ $\times$ 
8$^{\circ}$ region centered on M81. We then apply the 5$\sigma$ \hone mass threshold and velocity limits of our survey. 

For each major halo, we find the properties of sub-halos that would be detectable as \hone clouds in our observations.  We repeat this process rotating the cube three ways. The difference between the expected and observed number of clouds is striking. In the simulation, we find an average of 41 clouds for the WMAP3 simulation and 50 clouds for the WMAP5 simulation (ranging from 12-135 and 16-162 clouds, respectively) per major halo with sufficient mass to be detected in our observations. Recall that 9 clouds were detected in the corresponding region of our observations.

\begin{figure}
\centering
\begin{tabular}{cc}
\includegraphics[width=3in]{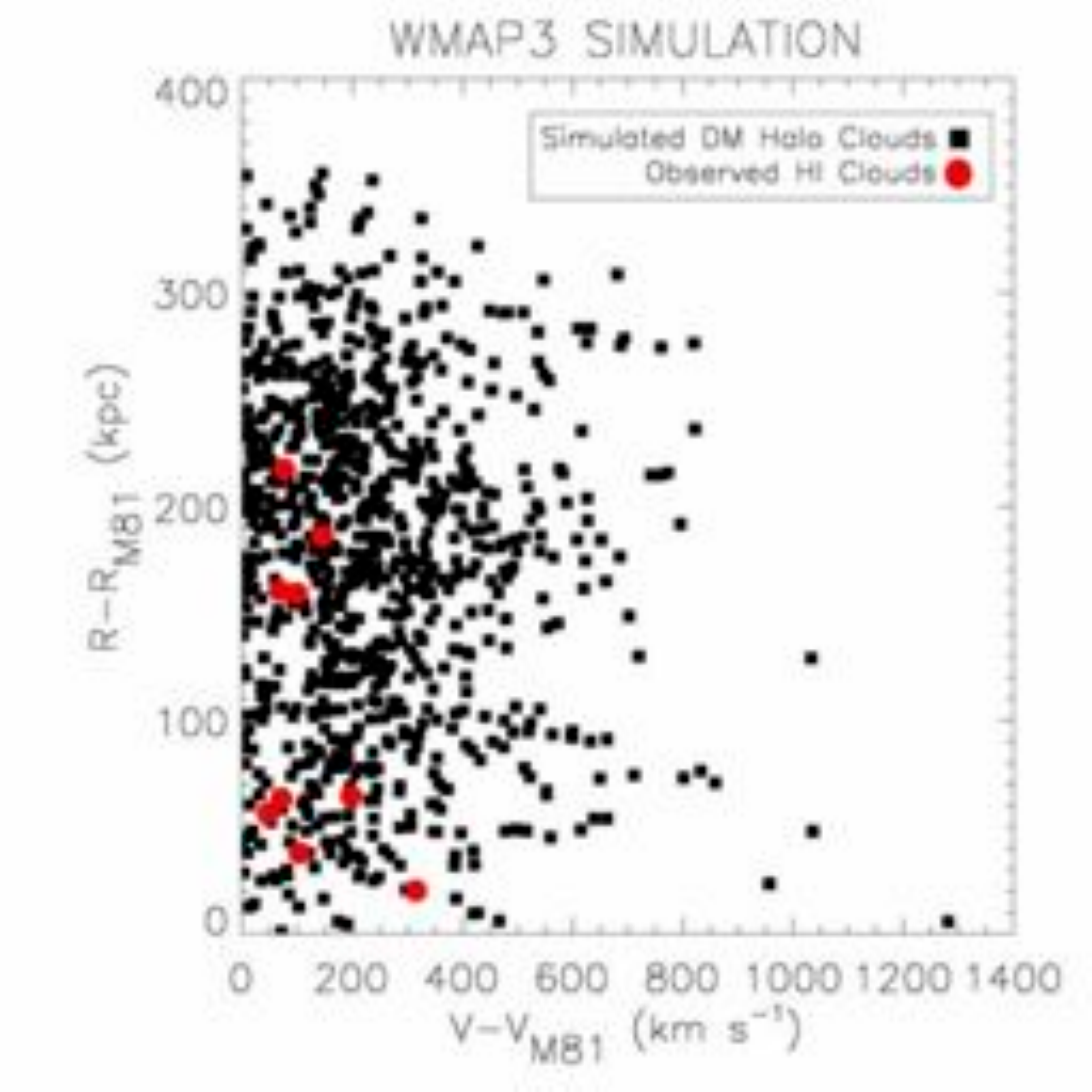} & 
\includegraphics[width=3in]{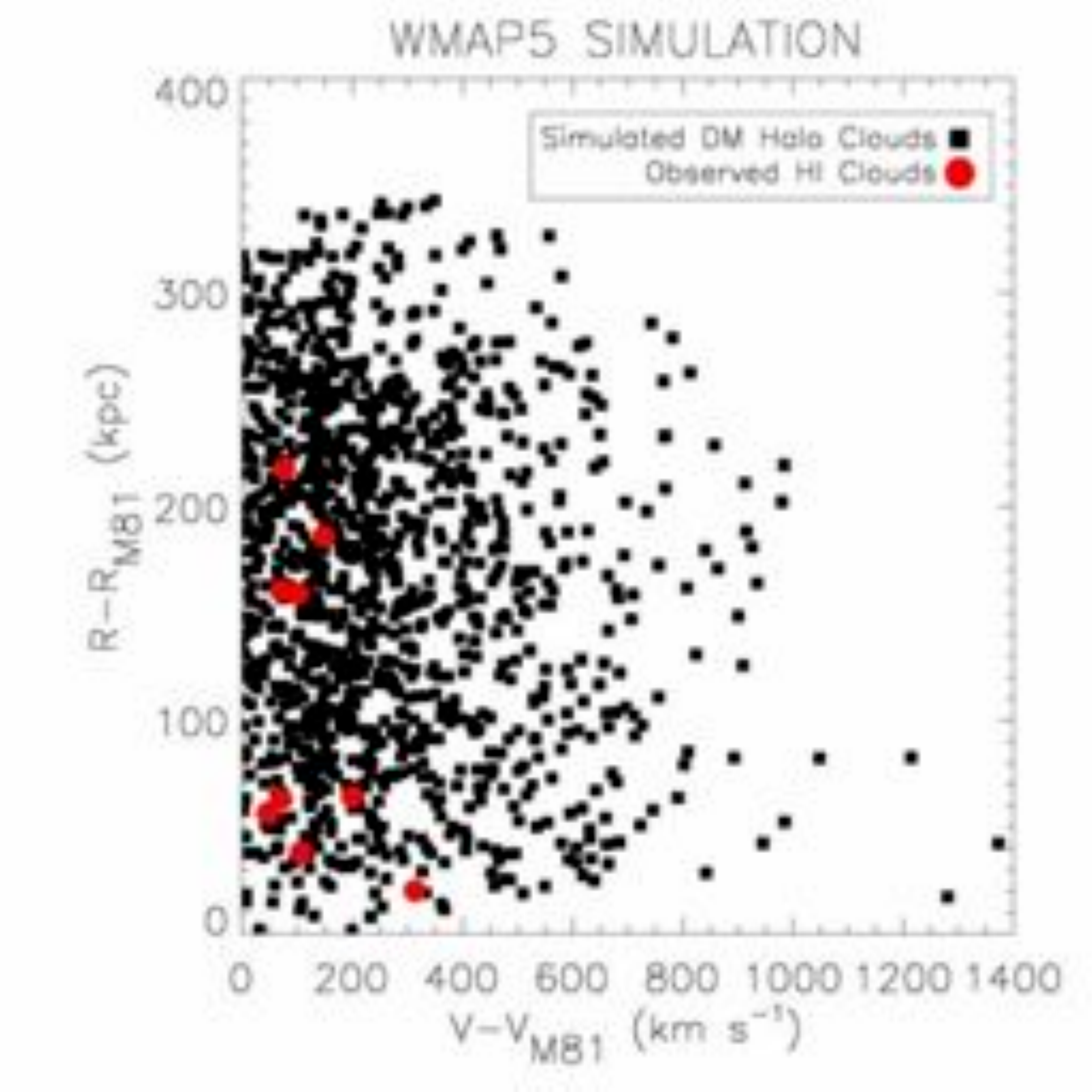} \\
\end{tabular}
\caption{Comparison of positions and velocites of \hone clouds in the 8$^{\circ}$ $\times$ 8$^{\circ}$ region centered on M81 with the expected properties of dark-matter halos with a  0.18\% \hone/total mass fraction. Left: Simulation with WMAP3 parameters. Right: Simulation with WMAP5 parameters. All detectable minihalos from each rotation of each major halo in the simulation are plotted. Note that the \hone clouds are much more clustered in position and velocity than dark-matter halos.}
\label{2dhistfig}
\end{figure}

We compare the positions and velocities of the detected clouds with the dark matter halos identified in the simulation, using a 2-dimensional Kolmogorov-Smirnov (KS) test. Figure \ref{2dhistfig} is a 2-dimensional phase-space diagram showing the distribution of positions and velocities of all observed \hone clouds, along with all detectable subhalos, around each major halo, including all rotations, in the simulation.    It is clear from the figure that the \hone clouds do not fall in the same region as the simulations; the \hone clouds are much more concentrated around the central galaxy in both position and velocity than the dark-matter halos. The results of the 2-dimensional KS-test bear this out, giving an average probability of 8.0 $\%$ for WMAP3 and 7.0 $\%$ for WMAP5 (across all rotations of all major halos) that the position-velocity distributions are consistent.

\section{Discussion, Conclusions and Future Work}
\label{consec}

We observed a wide area of the nearby M81 filament in an attempt to find \hone clouds tracing dark matter minihalos predicted by $\Lambda$CDM cosmological simulations. We detected 5 new \hone clouds, bringing the total to 13 \hone clouds in the M81 Filament including our previously published observations.

Of the 13 clouds detected in the observed region, 4 are likely to be part of the Milky Way HVC system. We compared the properties of the remaining 9 \hone clouds to two cosmological dark matter N-body simulations. We find that there are far fewer \hone clouds than simulations predict, and that the phase space distribution of the detected \hone clouds does not match that of the simulated clouds. There are two possible explanations for this discrepancy, both of which may be true. First, the \hone to dark matter mass fraction in minihalos may be less than 0.18$\%$. Secondly, the phase space distribution of dark matter minihalos predicted by $\Lambda$CDM cosmological simulations may be incorrect.  

Because our simulations are dark-matter only, none of the simulated clouds are tidal debris (which would contain no dark matter). With these simulations, we cannot predict the properties of \hone clouds representative of tidal debris, but we can confidently say that the clouds we have detected do NOT match the properties expected for dark-matter minihalos. The detected \hone clouds are also not compatible with cold accretion, since we do not detect clouds near non-interacting galaxies--cold accretion clouds should be detectable regardless of interactions. In fact, recent research suggests that cold accretion clouds should be \textit{more} prevalent in non-interacting groups, due to the lack of mitigating environmental effects \citep{ost10}.

Therefore, we conclude that the \hone clouds we have detected are not likely to be tracers of the predicted dark-matter minihalos in the M81 Filament. Instead, they are most likely generated through galaxy interactions. In order to understand the detailed role of galaxy interactions in generating \hone clouds, further simulations including gas physics and deeper observations of more filaments are needed.

This study allows us to constrain the number of
detected \hone clouds that may be self-gravitating. In a study of gravitationally bound molecular clouds, \citet{lar79,lar81} found a strong correlation between velocity dispersion and cloud size, given by 
	
\begin{equation}
\sigma_{v} \rm{(km~s^{-1})} = 1.1\rm{~L (pc)}^{0.38},
\end{equation}
	
\noindent where $L$ is the diameter of the cloud.  Our survey detection threshold of $\sigma_v$ =
10 km s$^{-1}$ gives a cloud size of 330 pc, so the angular resolution of our observations (10 kpc) is the
limiting factor in detecting self-gravitating \hone clouds. All of our clouds are unresolved in the
GBT beam. Our angular resolution limit gives a velocity dispersion of 36 km s$^{-1}$, so for any of
the detected clouds to be self-gravitating they must have a velocity dispersion less than or equal
to this value. Only two of the 9 clouds that we have determined to reside in the M81 group satisfy
this constraint. The implication of this constraint is that most interaction-generated \hone clouds in our
detection space are not self-gravitating.
	
On a side note, dark-matter filaments such as the M81 Filament are predicted to have diffuse \hone emission \citep{popping09}. The emission is predicted to be approximately 3 times fainter than our survey. Our observations do, however, place an upper limit on diffuse \hone emission from the cosmic web.

To estimate this, we determined our column density sensitivity using:

\begin{equation}
\left(\frac{\sigma_{N_{HI}}}{N_{HI}}\right)=1.82 \times 10^{18} \left(\frac{\sigma_{T_B}}{K}\right)
\left(\frac{\Delta V}{{\rm km}\hskip 1mm {\rm s}^{-1}}\right) \sqrt N
\end{equation}

The average 5$\sigma$ detection threshold for \hone column density was  4.4 $\times$ 10$^{18}$ \rm cm$^{-2}$. This column density falls within the ``\hone Desert", where the gas in the cosmic web is not affected by self-shielding and is therefore photoionized \citep{popping09}. Hence, we did not expect to detect diffuse \hone in the filament between the two galaxy groups. Our lowest 5$\sigma$ detection threshold for \hone column density was  $\sim$ 1 $\times$ 10$^{18}$  \rm cm$^{-2}$, which is approximately 3 times too high to detect the diffuse cosmic web in \hone emission. More sensitive GBT and EVLA observations of smaller regions, or surveys by future radio telescopes, could reach the sensitivity required to detect this emission.

\acknowledgments

KMC thanks the NRC Research Associateship program, the NRAO
Pre-Doctoral Fellowship program, and Vanderbilt University for funding support. 
{\it Facilities:} \facility{GBT}.

\bibliographystyle{aj}

\section*{Appendix A: Properties of Known Galaxies}

Our survey encompasses a large volume and we have detected a significant number of galaxies in \hone. We calculated \hone masses of all known galaxies in the observed volume using:

\begin{equation}
\left(\frac{M}{M_{\odot}}\right)=2.36 \times 10^5 \left(\frac{D}{{\rm Mpc}}\right)^2 \left(\frac{F}{{\rm Jy}}\right)
\left(\frac{\Delta V}{{\rm km}\hskip 1mm {\rm s}^{-1}}\right)
\end{equation}

\noindent Where $F$ is the total flux of the galaxy. 

Distances for galaxies in the M81 Filament were taken from \citet{kar07}. Distances to 
background galaxies were taken from NED where available; otherwise, they were calculated from the radial 
velocity, assuming a Hubble constant of H$_0$ = 75 km s$^{-1}$ Mpc$^{-1}$. Because 
of the high number of galaxies and continuum sources in the field, each galaxy spectrum
was re-baselined using a first-order fit to the surrounding spectral region, in order to calculate an
accurate \hone mass.

\clearpage
\onecolumn

\begin{table}
\begin{center}
\caption{Parameters of the Robert C. Byrd Green Bank Telescope System}
\begin{tabular}{ll}
\tableline\tableline
Telescope: \\
\hskip 3mm Diameter........................... & 100 m \\
\hskip 3mm Beamwidth (FWHM)................... &  9.1 \arcmin \\
\hskip 3mm Linear resolution ................. &  2.7 $D_{Mpc}$ kpc \\
Receiver: \\
\hskip 3mm Typical System Temperature ................ & 20 K \\
Spectrometer: \\
\hskip 3mm Bandwidth.......................... & 12.5-50 MHz\\
\hskip 3mm Resolution, Hanning Smoothed....... & 6.1 kHz (1.29 km s$^{-1}$) \\
\tableline
\end{tabular}
\end{center}
\end{table}

\begin{table}
\begin{center}
\caption{Observations Summary}
\begin{tabular}{ll}
\tableline\tableline
Center Frequency (MHz)		& 1419.4 \\
Bandwidth (MHz)			& 10.5 \\
Channel Width (kHz)			& 24.4 \\
Velocity Resolution (km s$^{-1}$)	& 5.2\\
Integration time (hours): 		& 187	\\
Typical RMS noise (mJy)		& 20\\
Sensitivity to \hone (5$\sigma$) & 3.2 $\times$ 10$^{6}$ M$_{\odot}$ \\
\tableline
\end{tabular}
\end{center}
\end{table}

\begin{table}
\begin{center}
\caption{M81 Filament \hone Cloud Properties\tablenotemark{a}}
\label{fil_cloudprop}
\begin{tabular}{l|c|c|c|c|c|c}
\hline\hline
Cloud \tablenotemark {b}	& Coordinate Designation  & $\Delta$D$_{M81}$ & T$_{peak}$	&V$_{LSR}$	&$\sigma_v$  & M$_{HI}$ ($\frac{D}{3.63 Mpc})^{-2}$ \\ 
		&                       & (kpc) &(K)	      & (km s$^{-1}$)		& (km s$^{-1}$) & $\times$ 10$^{6}$M$_{\odot}$\\ \hline
C08-1 & GBC J095007.2+695556  & 65 & 0.11 & 168 & 50 & 14.7 \\
C08-2 & GBC J100250.7+681949  & 63 & 0.10 & -102 & 55 & 22.5 \\
C08-3 & GBC J095244.9+681250  & 57 & 0.12 & 14 & 82 & 26.7 \\
C08-4 & GBC J100145.0+691631  & 38 & 0.30 & 74 & 28 & 83.7 \\
C08-5 & GBC J095506.3+692205 & 21 & 0.07 & 283 & 36 & 6.9 \\
C09-1 & GBC J073236.0+654302 & 900 & 0.06 & -147 & 20 & 1.8 \\
C09-2 & GBC J072528.5+654922 & 937 & 0.15 & -113& 20 & 7.5 \\
C09-3 & GBC J073612.1+673325 & 826 & 1.97 & -202 & 30 & 62.6 \\
C10-1   & GBC J092635.8+702850 & 186 & 0.13 &-178 & 57 &3.6  \\
C10-2   & GBC J101926.3+675222 & 161 & 0.20 & 38 & 26 & 3.4\\
C10-3   & GBC J091952.7+680937 & 218 & 0.17 &-108 & 42 & 5.1\\
C10-4  & GBC J1022:39.1+684057 & 159 & 0.30 & 69 & 68 & 12.0 \\
C10-5  & GBC J071051.4+654428 & 1024 & 0.19 & -301 & 47 & 2.8 \\ 
\hline
\tablenotetext{a}{Includes clouds discovered in \protect{\citet{chy08}} and \protect{\citet{chy09}}.}
\tablenotetext{b}{C08 indicates clouds from \protect{\citet{chy08}}, C09 indicates clouds from \protect{\citet{chy09}}, and C10 indicates clouds that are new in this paper.}
\end{tabular}
\end{center}
\end{table}

\begin{table}
\begin{center}
\caption{Properties of Known Galaxies\tablenotemark{a}}
\begin{tabular}{l|c|c|c|c|c}
\tableline\tableline
Name 		& $\alpha$\tablenotemark{b}	& $\delta$\tablenotemark{b} & D\tablenotemark{c}	&V$_{LSR}$\tablenotemark{d}	& M$_{HI}$ \\ 
		& (J2000)		  & (J2000)		     & Mpc		& km s$^{-1}$			&$\times$ 10$^{9}$M$_{\odot}$\\ \tableline
UGC 3580   	& 06:55:30.8 & 69:33:47 & 15.90 & 1193   & 2.59  \\
NGC 2366	& 07:28:55.7 & 69:12:59 & 3.19  &  102   & 0.68   \\
KUG 0724+695	& 07:29:50.0 & 69:25:26 & 12.02 & 902    & 0.27 \\
NGC 2403	& 07:36:52.7 & 65:35:52 & 3.30  &  134 &   3.38   \\
KKH 44  	         & 08:16:38.5 & 69:20:49 & 14.08 & 1056   & 0.06  \\
UGC 4305	& 08:19:06.5 & 70:43:01 & 3.39  &  160   & 0.91 \\
KDG 52 		& 08:23:56.2 & 71:01:36 & 3.55  &  116   & 0.02 \\
DDO 53		& 08:34:08.6 & 66:11:03 & 3.56  &  23    & 0.05   \\
UGC 4483	& 08:37:03.1 & 69:46:44 & 3.21  &  159   & 0.04  \\
NGC 2787  	& 09:19:18.5 & 69:12:12 &  9.31 & 716    & 0.72  \\
Holmberg I	& 09:40:32.3 & 71:10:56 & 3.84  & 152    & 0.16  \\
BK1N		& 09:45:15.3 & 69:23:22 &  7.86 & 584    & 0.06  \\
NGC 2976	& 09:47:15.6 & 67:54:49 & 3.56  &  6     & 0.52 \\
NGC 2985  	& 09:50:22.2 & 72:16:43 & 16.45 & 1213   & 3.17  \\
NGC 3027  	& 09:55:40.6 & 72:12:13 & 14.12 & 1061   & 4.74  \\
M81  		& 09:55:33.5 & 69:03:60 & 3.63  & -32    & 2.67 \\
M82		         & 09:55:53.9 & 69:40:57 & 3.53  &  206   & 0.75 \\
NGC 3077	& 10:03:21.0 & 68:44:02 & 3.82  &  17    & 1.01  \\ 
UGC 5423  	& 10:05:30.6 & 70:21:52 & 5.30  & 350    & 0.03  \\ 
HIJASS J1021+68	& 10:20:30.1 & 68:39:25 & 3.70  & 64     & 0.08 \\
UGC 5612  	& 10:24:06.5 & 70:52:56 & 13.67 & 1025   & 0.98 \\
IC 2574		& 10:29:19.0 & 68:27:42 & 4.02  & 79     & 1.36  \\
UGC 5692 	& 10:30:35.0 & 70:37:07 & 4.00  & 69     & 0.006 \\
NGC 3403  	& 10:53:54.8 & 73:41:25 & 15.35 & 1216   & 1.84  \\
\tableline
\tablenotetext{a}{Values for M81 and NGC 2403 group galaxies are taken from \protect{\citet{chy08}} and \protect{\citet{chy09}}.}
\tablenotetext{b}{From NED}
\tablenotetext{c}{From \protect{\citet{kar07}}, or cz/HO for D$\ge$ 7 Mpc}
\tablenotetext{d}{From our observations}
\end{tabular}
\end{center}
\end{table}

\begin{figure}
\centering
\begin{tabular}{ccc}
\includegraphics[width=2in]{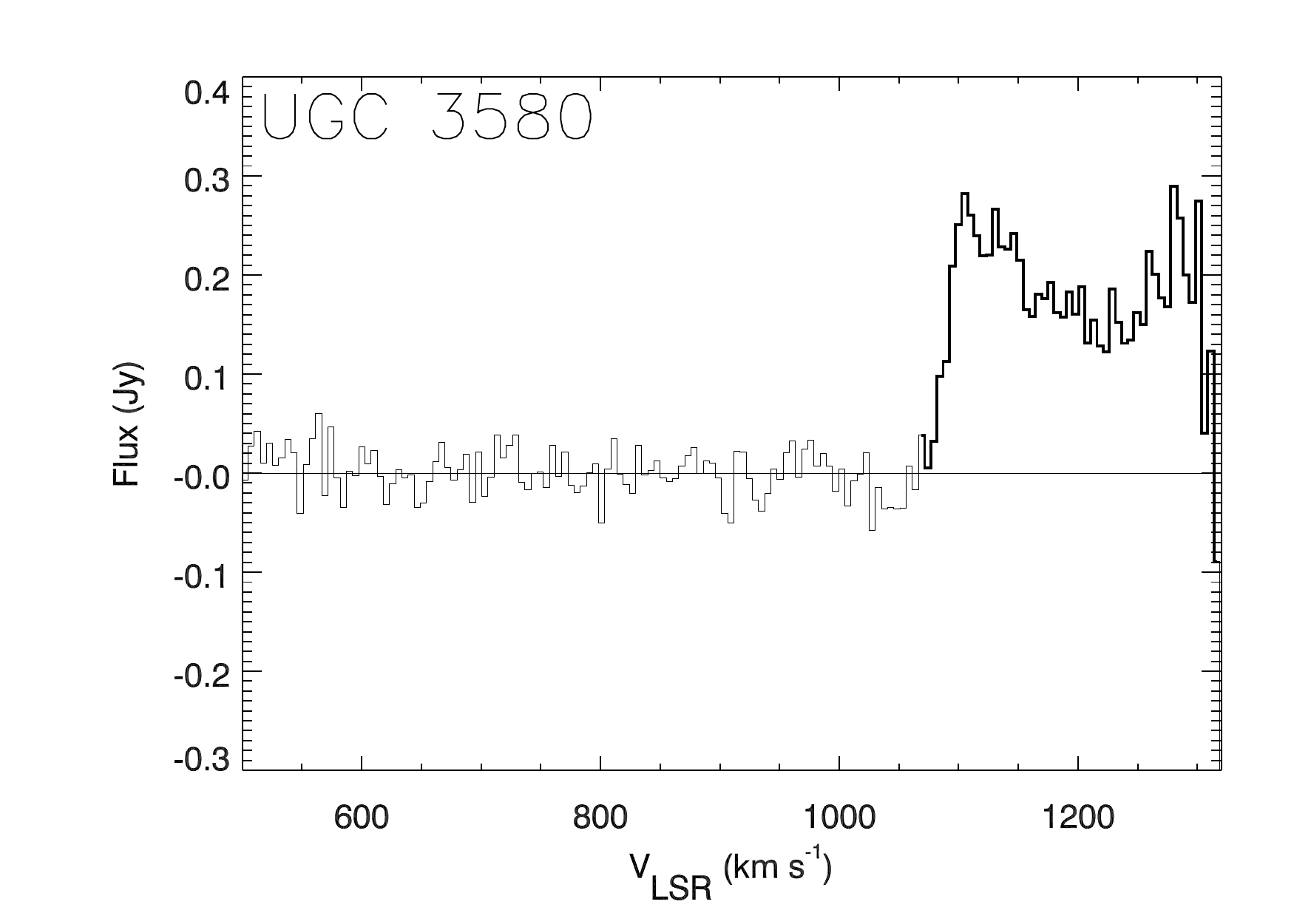} &\includegraphics[width=2in]{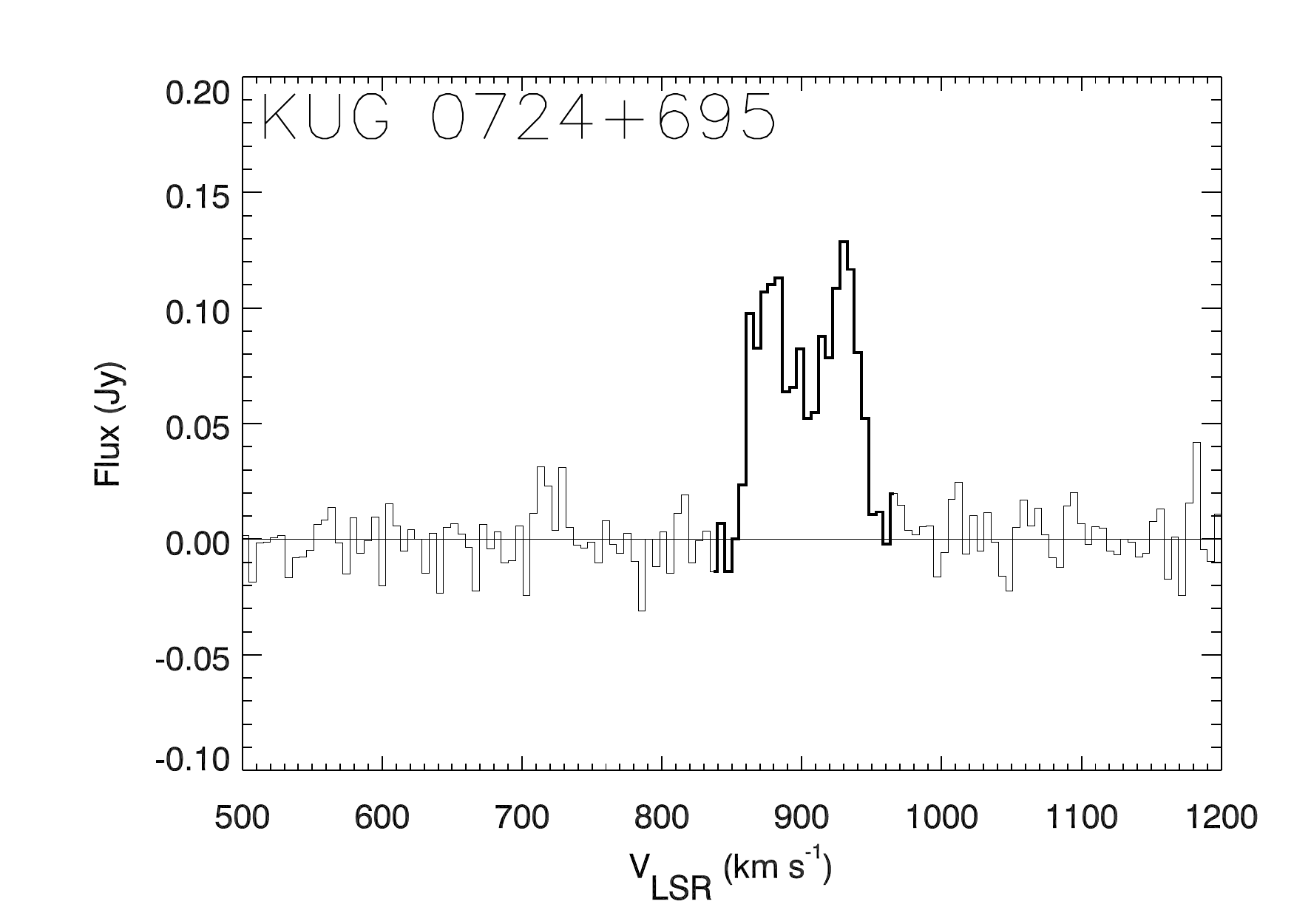} &  \includegraphics[width=2in]{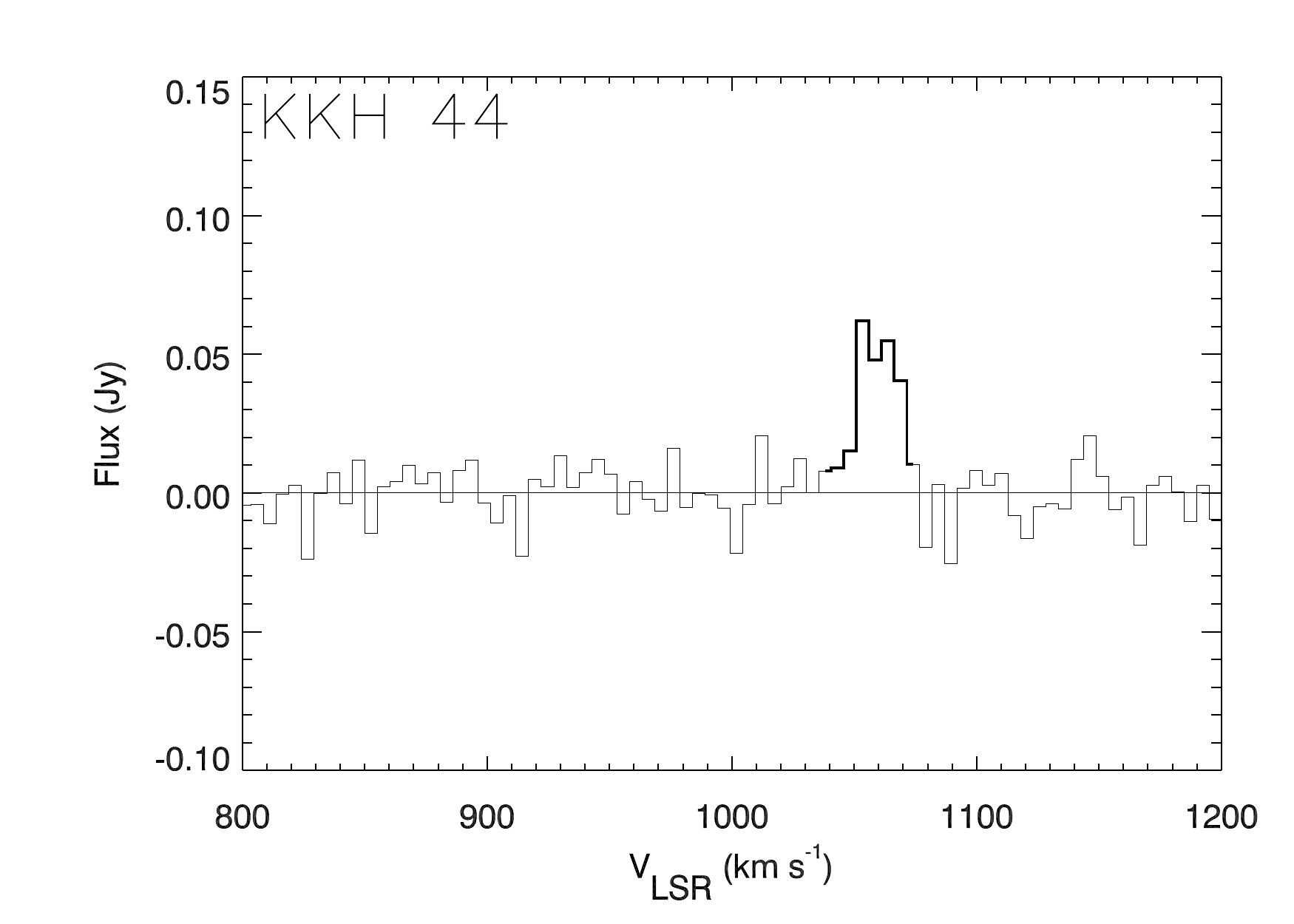} \\
\includegraphics[width=2in]{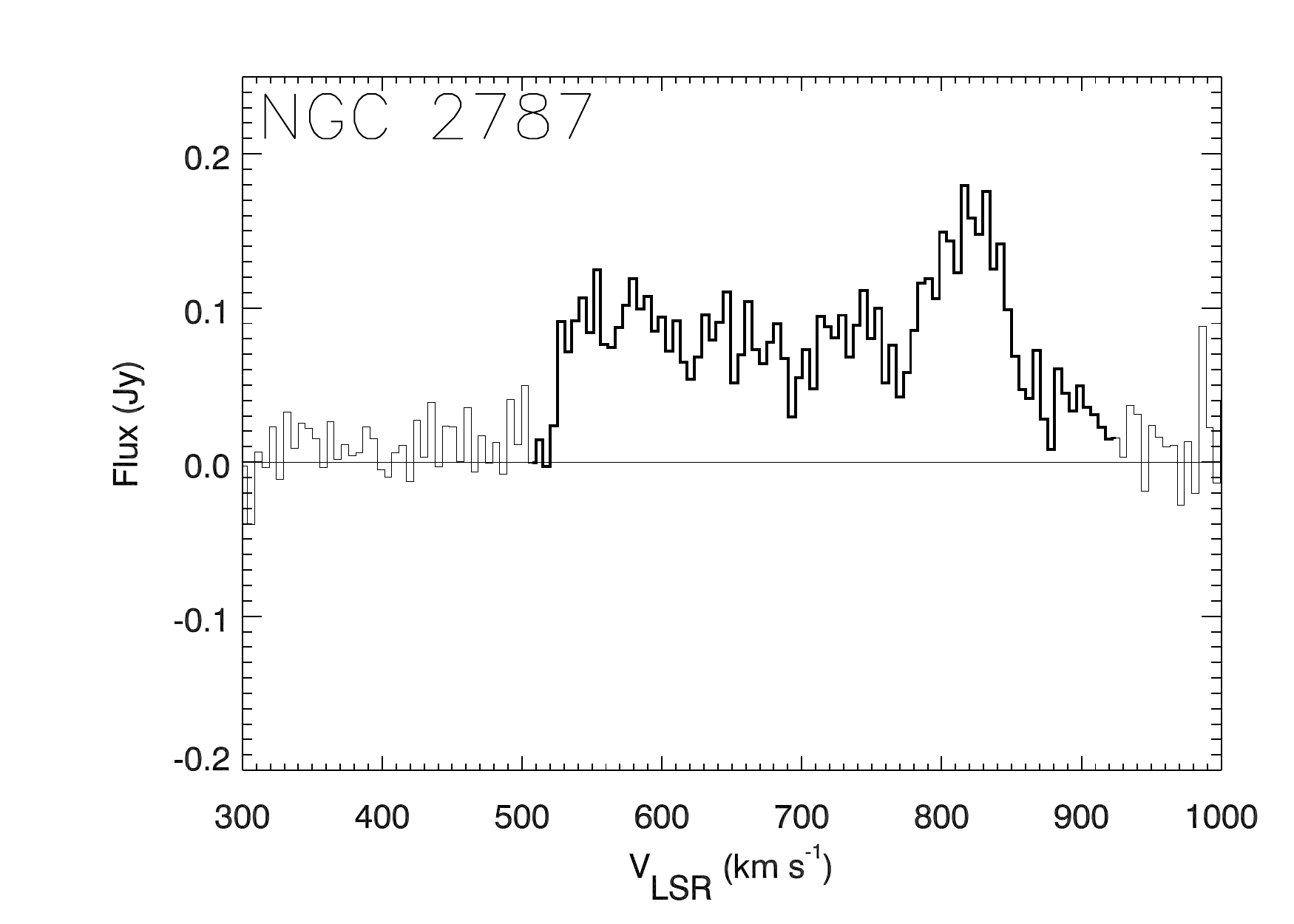} & \includegraphics[width=2in]{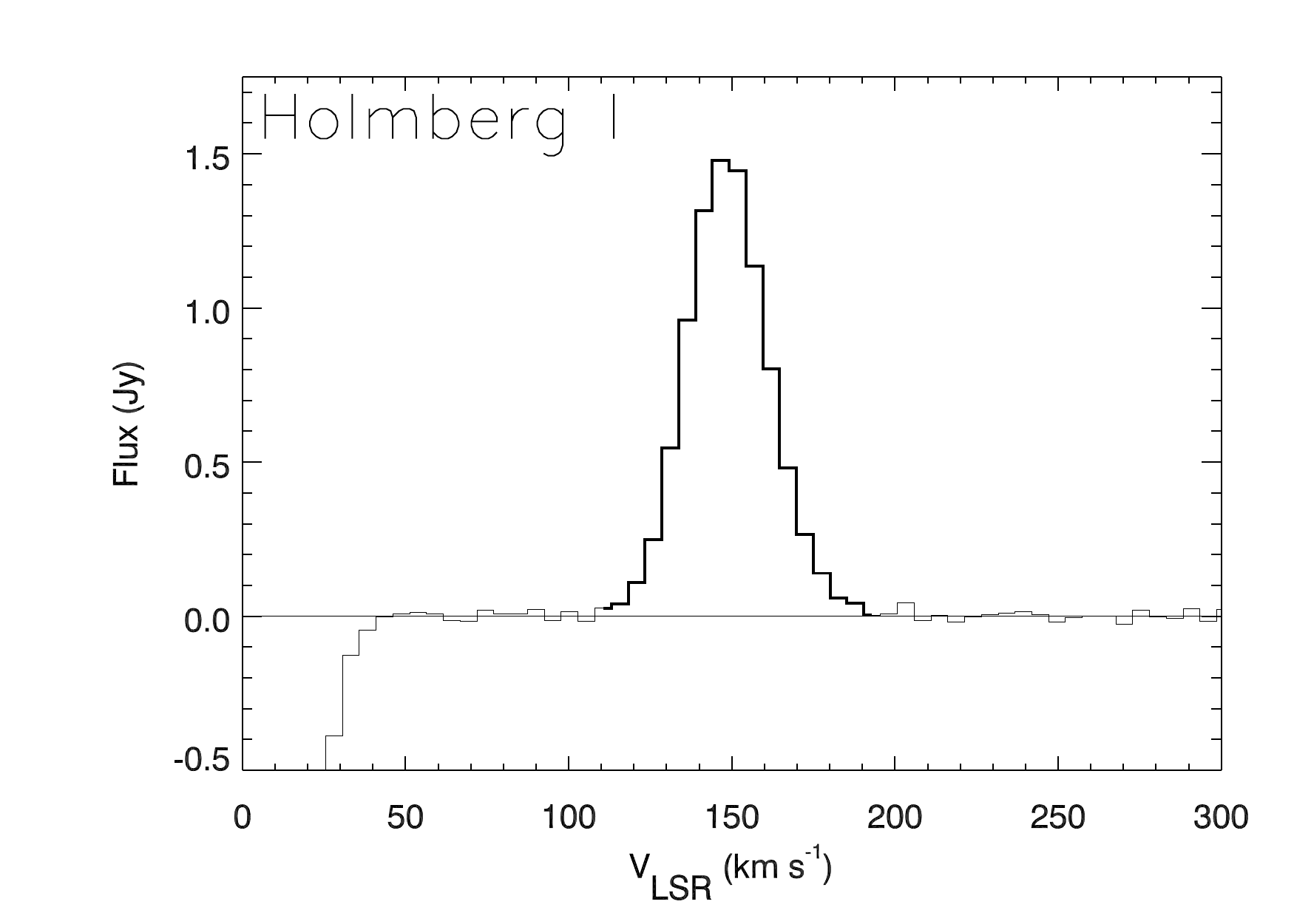}&\includegraphics[width=2in]{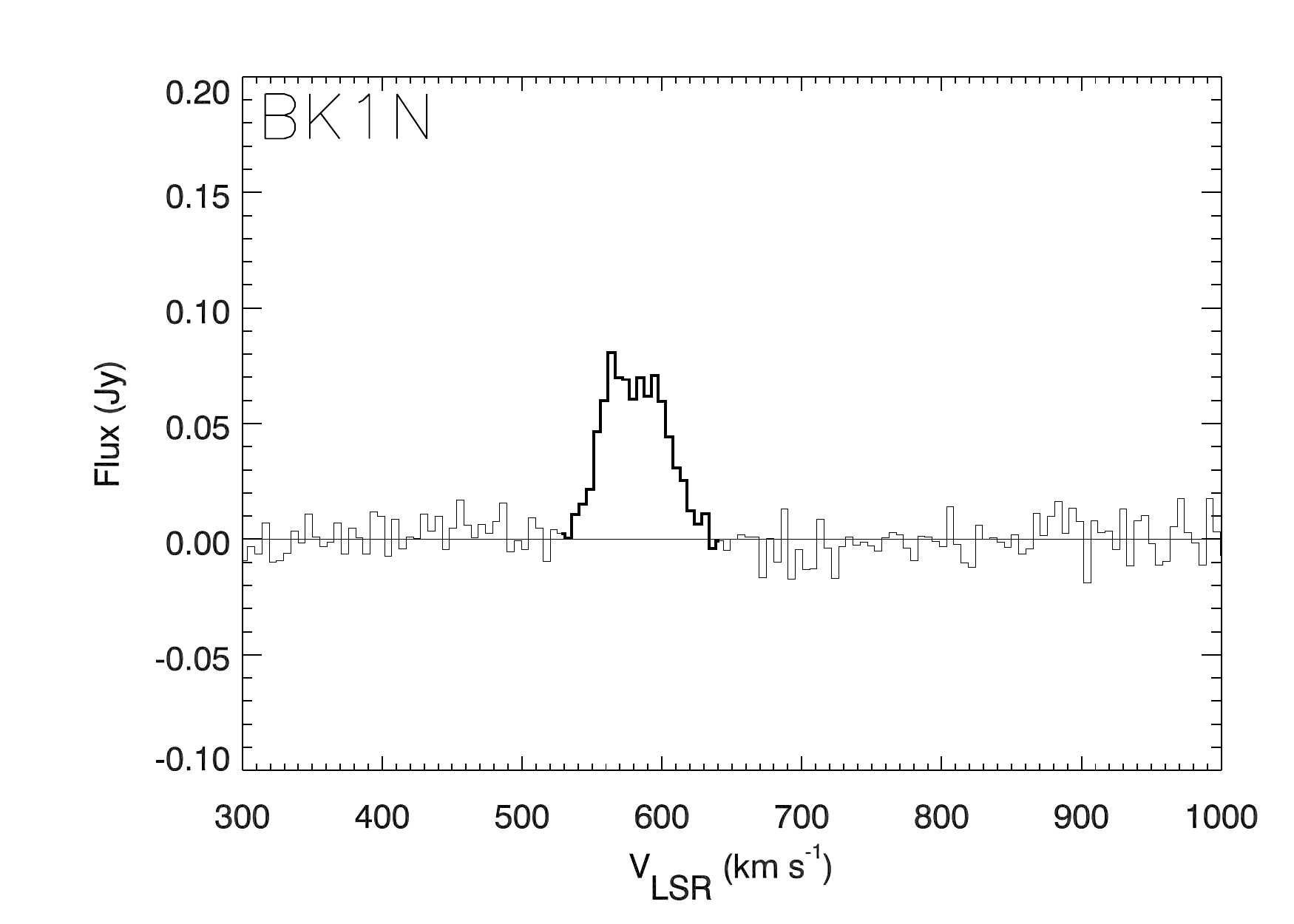} \\
\includegraphics[width=2in]{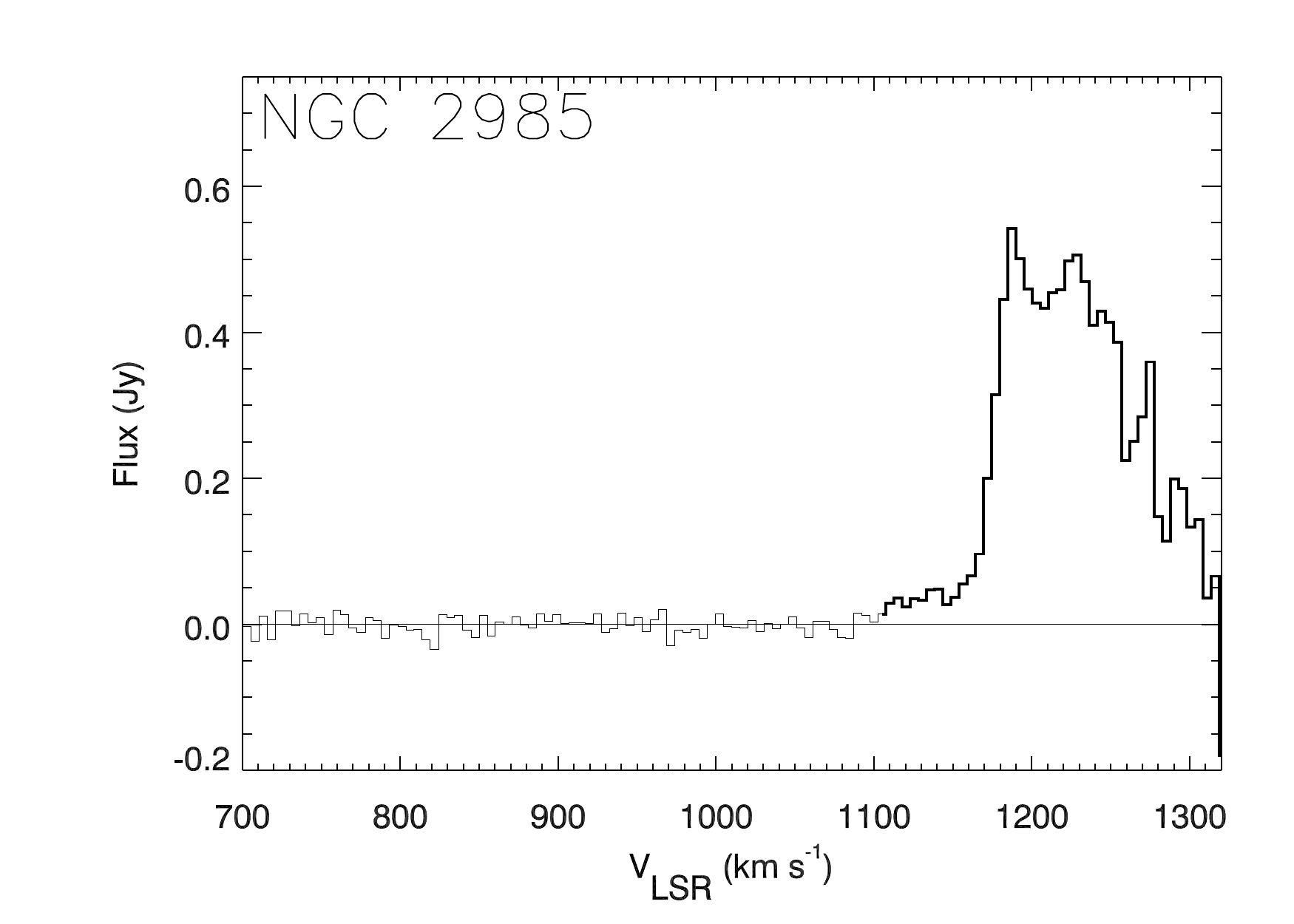}&\includegraphics[width=2in]{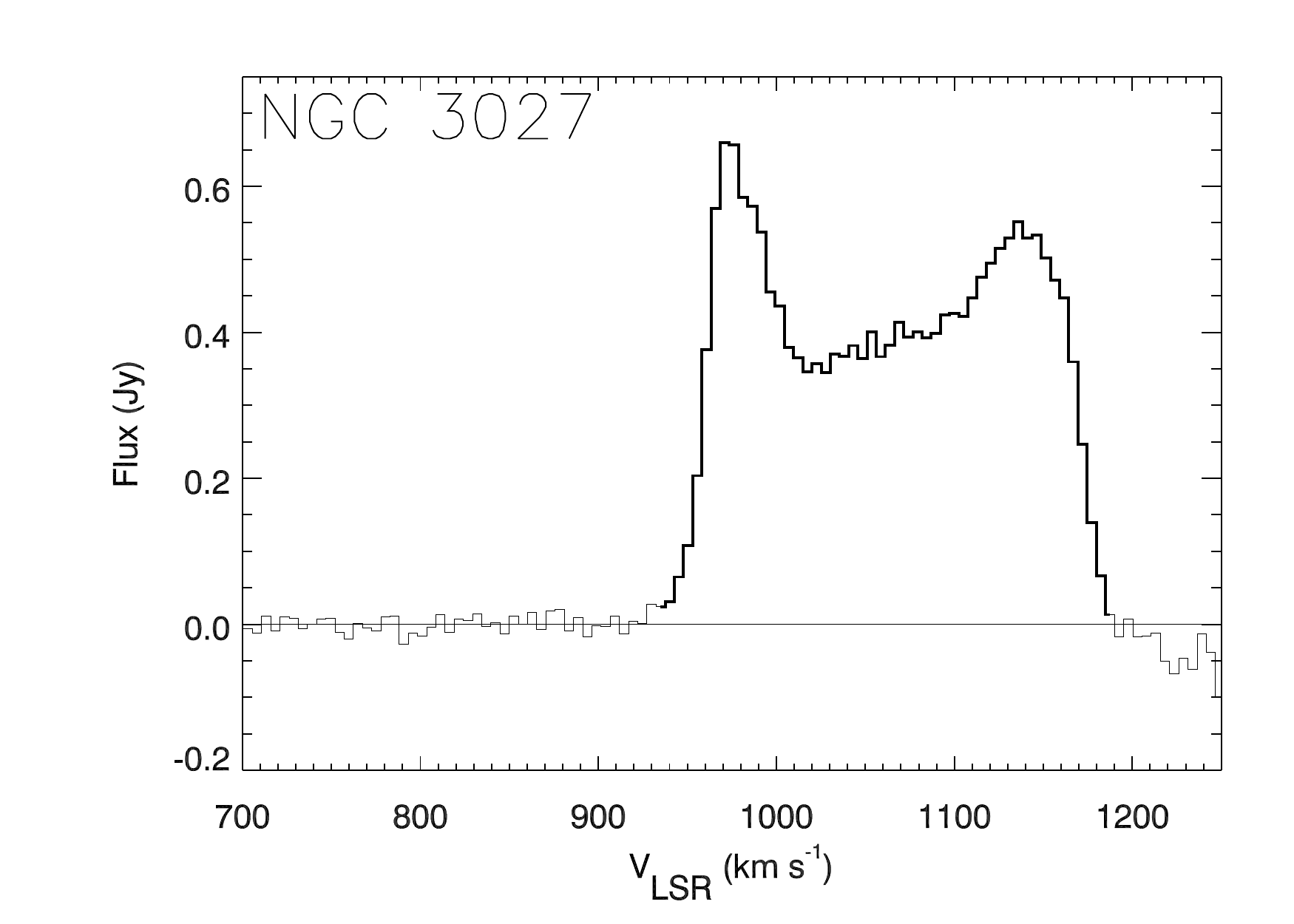} &\includegraphics[width=2in]{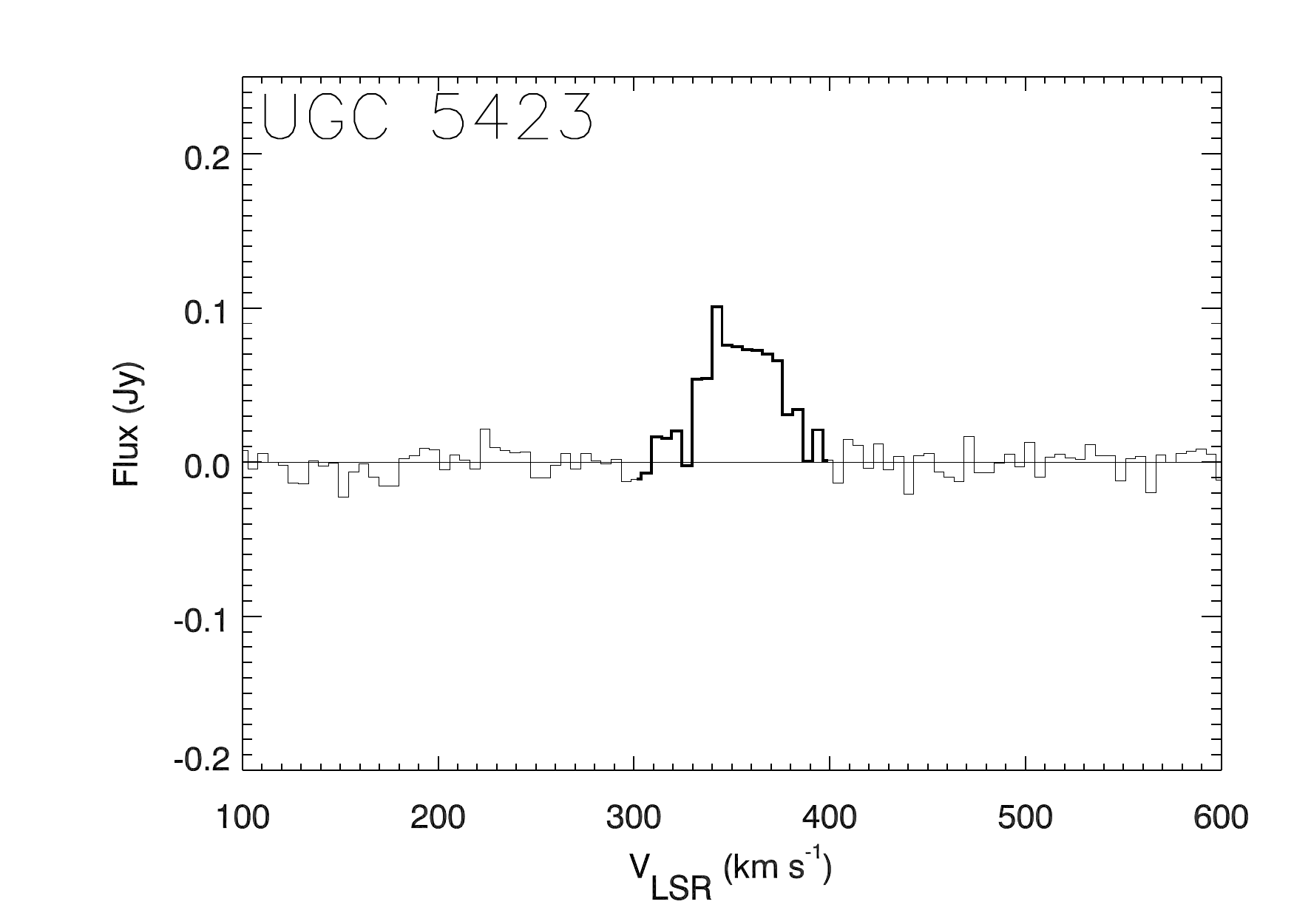}  \\
\includegraphics[width=2in]{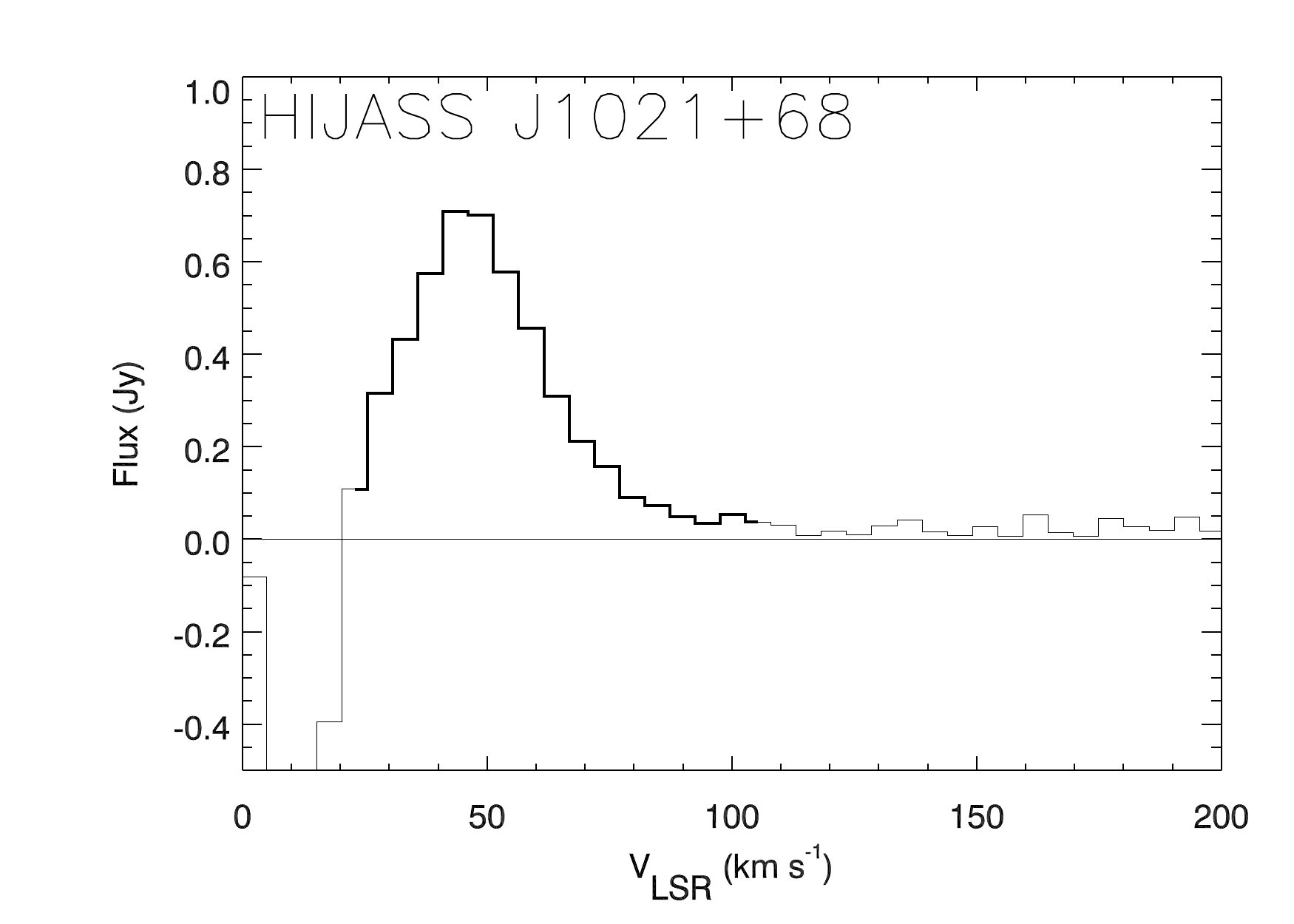}&\includegraphics[width=2in]{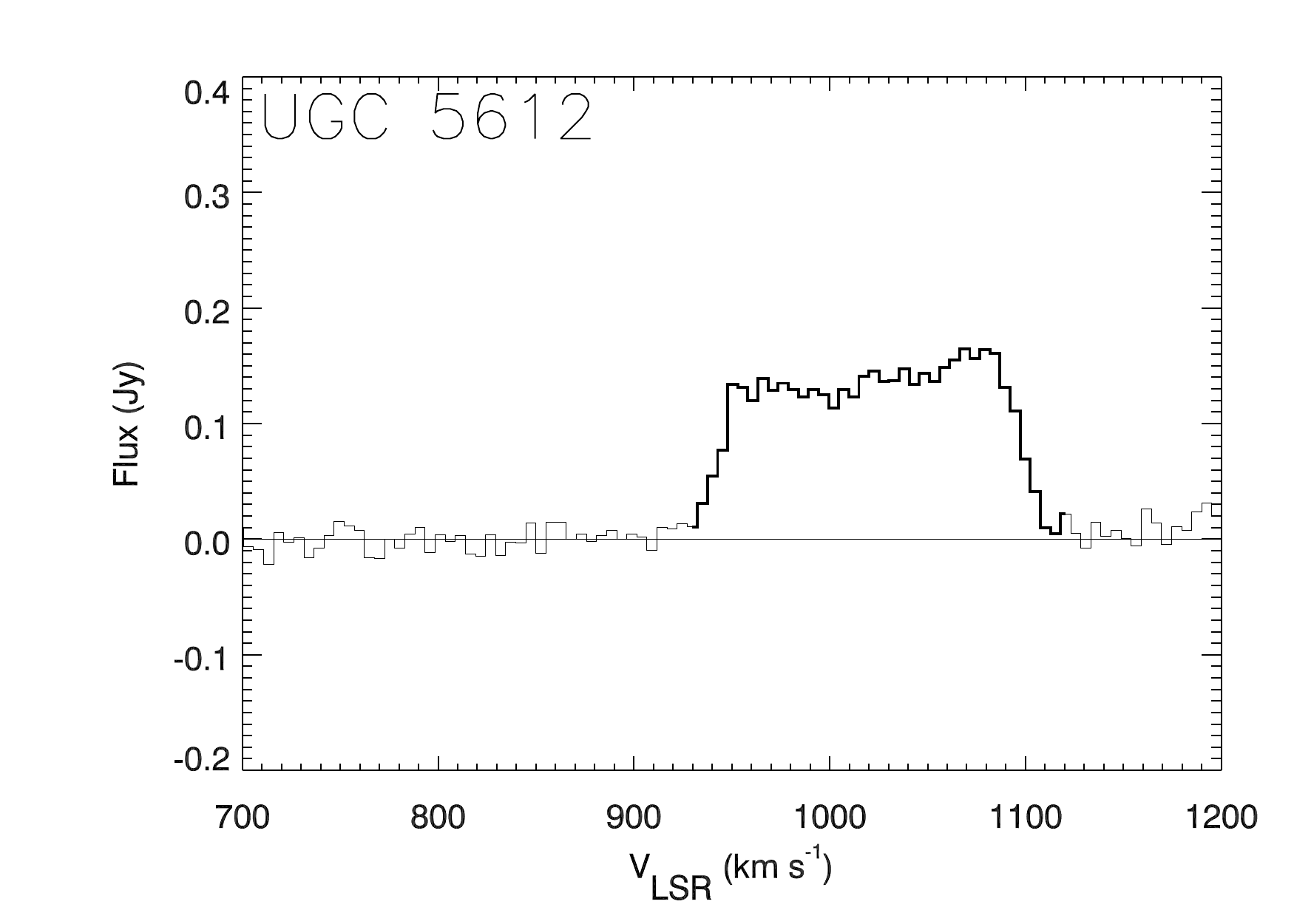} & \includegraphics[width=2in]{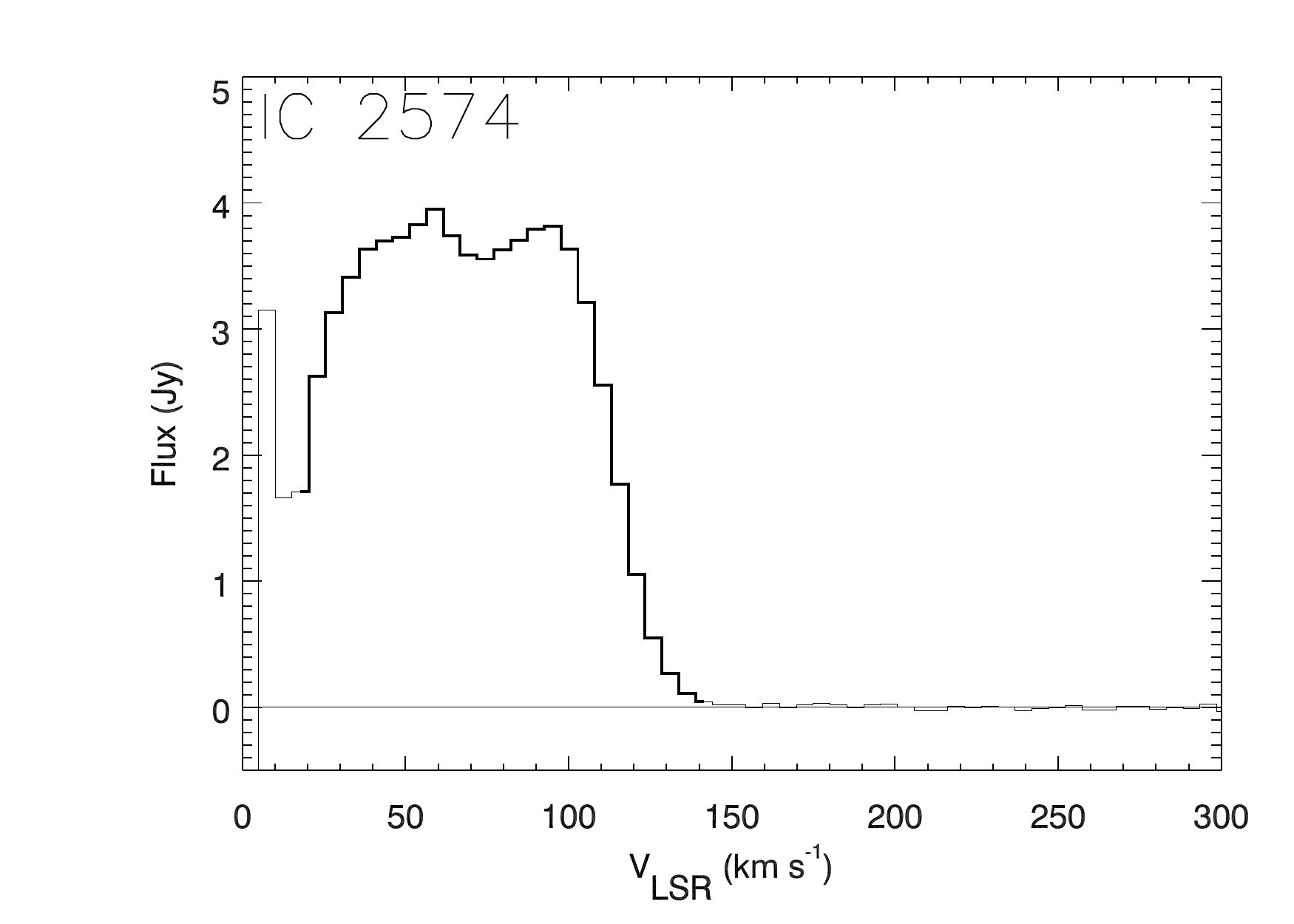}\\
\includegraphics[width=2in]{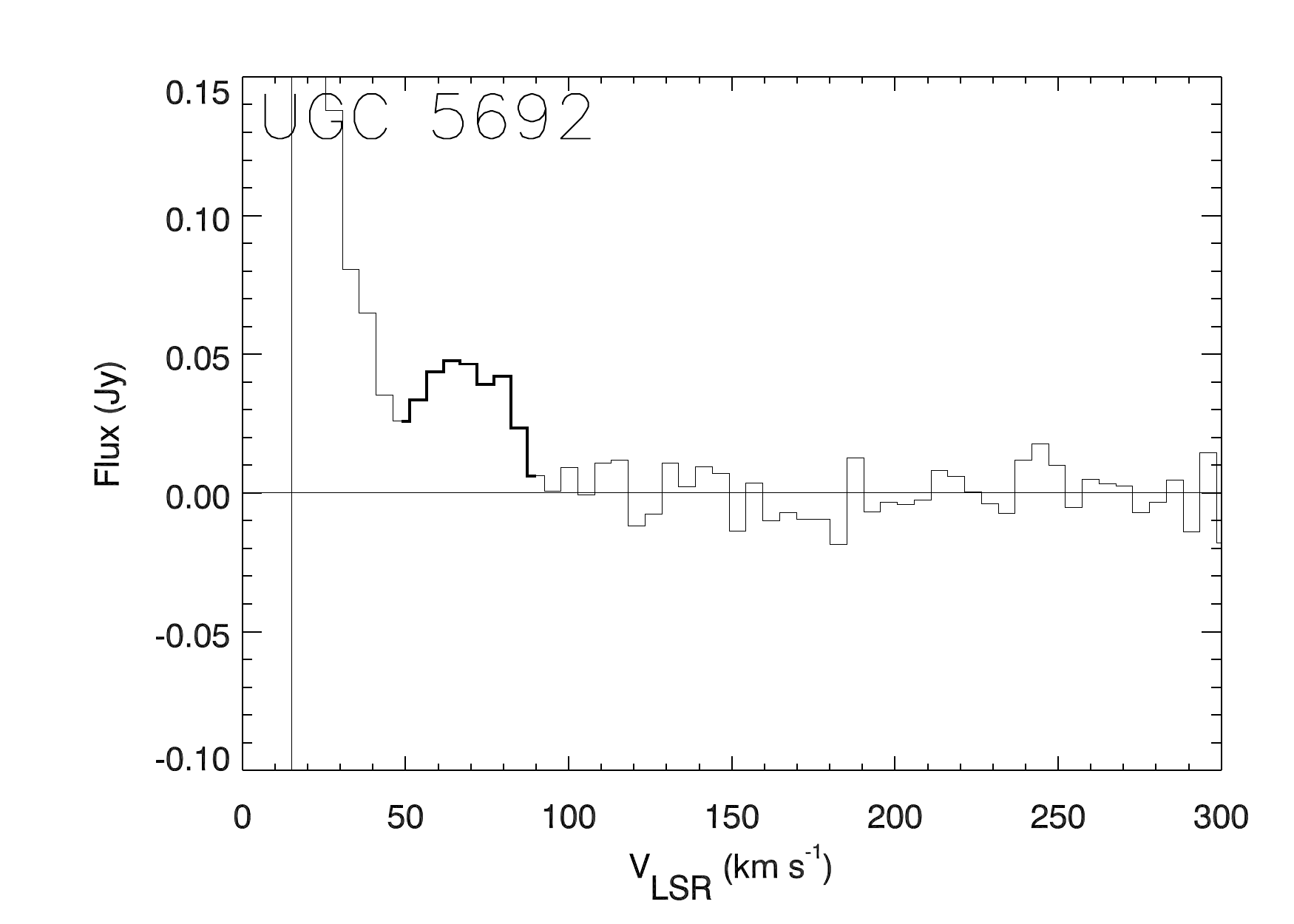}&\includegraphics[width=2in]{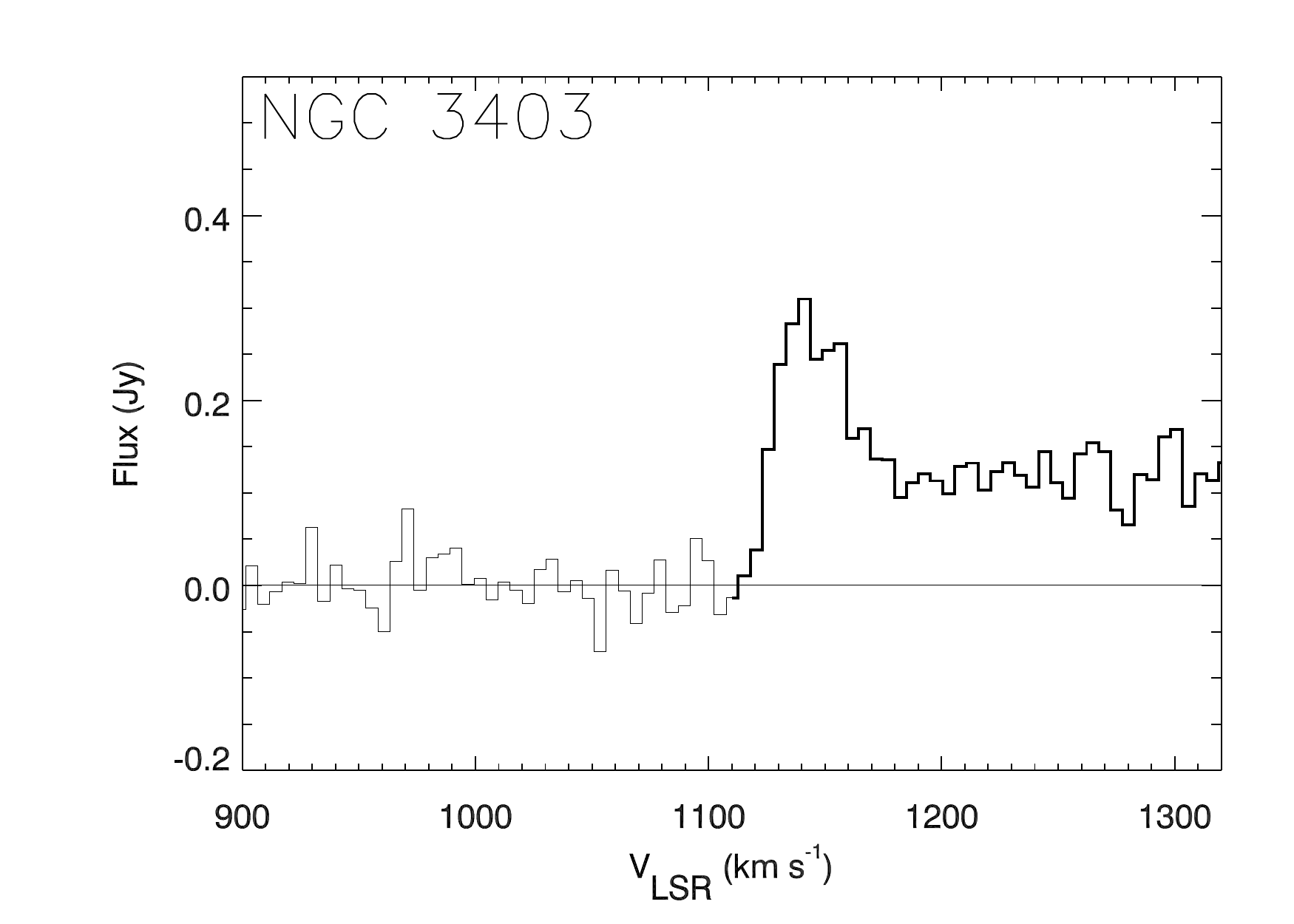} &\\
\end{tabular}
\caption{\hone line profiles of known galaxies within our observed field. Note that some galaxies do not fall completely within our bandwidth, so the derived mass will be low. Line profiles for the galaxies in the M81 and NGC 2403 groups  can be found in \protect{\citet{chy08}} and \protect{\citet{chy09}}.}
\label{glx_ispec}
\end{figure}

\end{document}